\documentstyle[preprint,aps,amssymb]{revtex}
\tightenlines
\begin{document}
%\draft
\title{Identification and Control of Symmetric Systems}
\author{R. O. Grigoriev}
\address{James Franck Institute, University of Chicago, Chicago, Illinois
60637}
\date{\today}
\maketitle

\begin{abstract}
In a recent paper [Phys. Rev. E {\bf 57}, 1550 (1998)] we demonstrated that the
symmetries of the evolution equation and the target state have a profound
effect on the selection of the admissible control parameters. In the present
paper we extend these results to the case of time-periodic target trajectories
and inexact symmetries. We also argue that the problem of phase space
reconstruction is affected by the presence of symmetries similarly to the
control problem.
\end{abstract}

%%%%%%%%%%%%%%%%%%%%%%%%%%%%%%%%%%%%%%%%%%%%%%%%%%%%%%%%%%%%%%%%%%%%%%%%%%%%%%%
\section{Introduction}
\label{s_intro}
%%%%%%%%%%%%%%%%%%%%%%%%%%%%%%%%%%%%%%%%%%%%%%%%%%%%%%%%%%%%%%%%%%%%%%%%%%%%%%%
 
The desire to improve performance of many practically important systems and
devices often calls for shifting their operating range into a highly nonlinear
area, which after a series of bifurcations usually leads to irregular chaotic
behavior. This kind of behavior, however, is rarely desired, while substantial
benefits could be obtained by making the dynamics regular. This goal can
typically be achieved by applying small preprogrammed perturbations to steer
the system towards a periodic orbit with desired properties, which is broadly
referred to as chaos control.

The goal of the present paper is to highlight some of the problems that arise
in controlling systems which possess some kind of symmetry. Even though the
importance of symmetries in chaotic dynamics has been recognized by a number of
authors \cite{chossat,golubitsky}, symmetric systems did not receive adequate
treatment in the general framework of chaos control primarily because the
question of symmetry is largely ignored by the theory of deterministic chaos as
well as data analysis and control theory. All three disciplines regard
symmetric systems as nongeneric and, therefore, not very interesting and
important. 

However, many practically important dynamical systems, such as spatially
extended chaotic ones, are symmetric, and thus cannot be successfully treated
using the formalism developed for generic systems. Indeed, such phenomena as
fluid flows, convection or chemical reactions often take place inside symmetric
containers --- cylinders, spheres, pipes and annuli. As a result, the dynamical
equations also show rotational and translational symmetries. Even the dynamics
of unbounded systems is often significantly influenced by the symmetries of the
physical space. Although the presence of symmetries usually simplifies the
analysis of the dynamics, it also makes system identification and control more
complicated due to the inherent degeneracies of the evolution operators. In
fact, the presence of symmetries, explicit or implicit, makes a number of
single-control-parameter methods fail \cite{ding,romeiras,petrov_1}, calling
for multi-parameter control \cite{warncke,self_prl,barreto,locher}.

In order to see how the control problem is affected by symmetries, we consider
(following the analysis conducted in \cite{self_pre}) a general discrete-time
system (the arguments for continuous-time systems are very similar), whose
evolution is described by the map ${\bf F}:\,{\mathbb R}^{n_x}\times {\mathbb
R}^{n_u} \rightarrow {\mathbb R}^{n_x}$, such that
 \begin{equation}
 \label{eq_nl_id}
 {\bf x}^{t+1}={\bf F}({\bf x}^t,{\bf u}),
 \end{equation}
where ${\bf x}^t$ is the $n_x$-dimensional state of the system and ${\bf u}$
is the $n_u$-dimensional vector of system parameters. The objective of control
is to make the system follow the (possibly unstable) periodic target trajectory
$\bar{\bf x}^t$. Let us linearize equation (\ref{eq_nl_id}) about this target
trajectory to obtain
 \begin{equation}
 \label{eq_lin_id}
 \Delta{\bf x}^{t+1}=A^t\Delta{\bf x}^t+B^t\Delta{\bf u}^t,
 \end{equation}
where
 \begin{equation}
 A^t={\bf D}_{\bf x}{\bf F}(\bar{\bf x}^t,\bar{\bf u})
 \end{equation}
is the Jacobian matrix, which determines the stability properties of the target
trajectory, and 
 \begin{equation}
 B^t={\bf D}_{\bf u}{\bf F}(\bar{\bf x}^t,\bar{\bf u})
 \end{equation}
is the control matrix, which defines the linear response of the system to
perturbation of system parameters. According to linear system theory
\cite{rubio}, if the target trajectory $\bar{\bf x}^t$ is unstable, it can be
stabilized by an appropriate feedback through the time-dependent control
perturbation $\Delta{\bf u}^t$, provided the matrices $A^t$ and $B^t$ satisfy
certain conditions. In the present study we concentrate on selecting from the
complete set of available {\em system} parameters a minimal set of {\em
control} parameters, whose perturbation allows the stabilization of the target
state, i.e., on making an appropriate choice of the control matrix $B^t$, given
the Jacobian $A^t$. We will see below that the constraints affecting the choice
of control parameters can be easily obtained from the symmetry properties of
the system and the controlled state. What is more interesting, symmetry allows
one to determine the minimal number of control parameters even when the
Jacobian $A^t$ describing the local dynamics is unknown.

Discrete-time evolution equations of type (\ref{eq_nl_id}) are often obtained
as a result of phase space reconstruction of a continuous-time system when the
dynamical equations
 \begin{equation}
 \label{eq_nl_id_cont}
 \dot{\bf s}(t)={\bf\Phi}({\bf s}(t),{\bf u})
 \end{equation}
describing its evolution are unknown. Here ${\bf s}(t)\in{\mathcal Q}$ denotes
the $n_s$-dimensional internal state of the system, and ${\bf \Phi}$ is an
unknown vector field on the phase space manifold ${\mathcal Q}$. Generically,
such reconstruction is possible when the measurement of a single scalar
time-dependent signal $y(t)$, which is a function of the system state ${\bf
s}(t)$, is available. Many practically interesting systems, symmetric ones in
particular, are, however, extremely nongeneric and require a number of
independent scalar signals for the complete reconstruction. Eckmann and Ruelle
\cite{eckmann} acknowledged that the choice of signals has to be made carefully
by trial and error. Certain general rules concerning this choice, however, can
be established on purely theoretical grounds, since this problem too can be
effectively treated based on the knowledge of underlying symmetries
\cite{king}.

The outline of the paper is as follows. In section \ref{s_steady} we discuss
the implications of symmetry for control of discrete-time linear systems in the
vicinity of steady target states. The results are generalized for time-periodic
target trajectories in section \ref{s_ltv}, and for continuous-time systems in
section \ref{s_contin}. Section \ref{s_violat} is devoted to the problem of
weak symmetry violation. The issue of phase space reconstruction is considered
in section \ref{s_recons}. The theoretic results are illustrated using several
simple examples in section \ref{s_examples}. Finally, we present our
conclusions in section \ref{s_summary}.

%%%%%%%%%%%%%%%%%%%%%%%%%%%%%%%%%%%%%%%%%%%%%%%%%%%%%%%%%%%%%%%%%%%%%%%%%%%%%%%
\section{Time-Invariant States}
\label{s_steady}
%%%%%%%%%%%%%%%%%%%%%%%%%%%%%%%%%%%%%%%%%%%%%%%%%%%%%%%%%%%%%%%%%%%%%%%%%%%%%%%

%%%%%%%%%%%%%%%%%%%%%%%%%%%%%%%%%%%%%%%%%%%%%%%%%%%%%%%%%%%%%%%%%%%%%%%%%%%%%%%
\subsection{Stabilizability and Controllability}
\label{ss_stab_cont}
%%%%%%%%%%%%%%%%%%%%%%%%%%%%%%%%%%%%%%%%%%%%%%%%%%%%%%%%%%%%%%%%%%%%%%%%%%%%%%%

Although our analysis is applicable to time-varying systems, we start for
simplicity by assuming that the target state is time-invariant, $\bar{\bf x}^t
=\bar{\bf x}$. Then the matrices $A^t$ and $B^t$ become constant, and we can
drop the time index in (\ref{eq_lin_id}) to obtain
 \begin{equation}
 \label{eq_lin_id_const}
 \Delta{\bf x}^{t+1}=A\Delta{\bf x}^t+B\Delta{\bf u}^t.
 \end{equation}
It is useful to introduce and compare two characterizations of the linearized
evolution equation (\ref{eq_lin_id_const}), which extremely simplify the
analysis of feedback control algorithms: {\em stabilizability} and {\em
controllability}.

The dynamical system (\ref{eq_lin_id_const}) or the pair $(A,B)$ is said to be
{\em stabilizable}, if there exists a {\em state feedback} 
 \begin{equation}
 \label{eq_feed_const}
 \Delta{\bf u}^t=-K\Delta{\bf x}^t,
 \end{equation}
making the system (\ref{eq_lin_id_const}) stable, i.e., it is possible to find
a {\em feedback gain matrix} $K$, such that all eigenvalues $\lambda'_k$ of the
matrix $A'=A-BK$ lie within a unit circle of the complex plane,
$|\lambda'_k|<1$, $\forall k$. Otherwise the system or the pair $(A,B)$ is
called {\em unstabilizable}. Indeed, substituting the feedback
(\ref{eq_feed_const}) into (\ref{eq_lin_id_const}) one obtains the linearized
evolution equation for the {\em closed-loop} system
 \begin{equation}
 \label{eq_lin_cloop_id}
 \Delta{\bf x}^{t+1}=(A-BK)\Delta{\bf x}^t,
 \end{equation}
with $\Delta{\bf x}={\bf 0}$ becoming the stable fixed point of the map
(\ref{eq_lin_cloop_id}), if and only if $A-BK$ is stable.

Since the magnitude of the control perturbation $\Delta{\bf u}^t$ is
proportional to the deviation $\Delta{\bf x}^t$ of the system from the target
state, feedback of the form (\ref{eq_feed_const}) is often called {\em
proportional} in the physics literature, although there are a number of other
terms used to denote this type of feedback. Control theory uses the term {\em
state} feedback to refer to the fact that the state of the system is used to
determine the control perturbation. At first sight equation
(\ref{eq_feed_const}) seems to impose strict limitations on the allowed form of
the feedback law. However, this is precisely the form demanded by a number of
widely used control algorithms \cite{ding,romeiras,rhode}.

Stabilizability is a property, which usually sensitively depends on the values
of system parameters. In the majority of practical applications, however, it is
preferable to have an adaptive control that would stabilize a given steady
state $\bar{\bf x}(\bar{\bf u})$ for arbitrary values of system parameters.
This is especially important, if one is to track the trajectory $\bar{\bf x}$
as parameters slowly vary, which might be advantageous in many applications,
e.g., for moving the operating point of a nonlinear device across a
bifurcation, from the stable region to the chaotic region. Such a control
scheme can be obtained, if the more restrictive condition of controllability,
which is essentially parameter-independent, is imposed on the matrices $A$ and
$B$. On the other hand, it can be demonstrated \cite{rubio} that the
controllability condition guarantees that the eigenvalues of the matrix $A-BK$
can be freely assigned (with complex ones in conjugate pairs) by an appropriate
choice of the matrix $K$. Therefore, if the system is controllable, it is
stabilizable as well, and by requiring controllability we satisfy both
conditions at once.

The $n_x$-dimensional linear system (\ref{eq_lin_id_const}) or the pair $(A,B)$
is said to be {\em controllable} if, for any initial state $\Delta{\bf
x}^{t_i}= \Delta{\bf x}_i$, times $t_f-t_i\ge n_x$, and final state $\Delta{\bf
x}_f$, there exists a sequence of control perturbations $\Delta{\bf
u}^{t_i},\cdots, \Delta{\bf u}^{t_f-1}$ such that the solution of equation
(\ref{eq_lin_id_const}) satisfies $\Delta{\bf x}^{t_f}=\Delta{\bf x}_f$.
Otherwise, the system or the pair $(A,B)$ is called {\em uncontrollable}.

The controllability condition can be represented in a number of different
equivalent forms. To obtain one particularly convenient form, we make the
trivial observation that, if it is possible to drive the linear system from an
arbitrary initial state $\Delta{\bf x}_i$ to an arbitrary final state
$\Delta{\bf x}_f$ in $n_x$ steps, it is possible to do the same in any number
of steps $n$ exceeding $n_x$. Suppose we let the system evolve under control
for $n_x$ steps from the initial state $\Delta{\bf x}^t$. The final state will
be given by\footnote{Here and below in the text we use the notation $(A)^n$ to
indicate that $A$ is taken to the power of $n$ to differentiate it from the
notation $A^t$, where index $t$ defines the time dependence.}
 \begin{equation}
 \label{eq_iter_gen}
 \Delta{\bf x}^{t+n_x}=
 (A)^{n_x}\Delta{\bf x}^t+\sum_{k=1}^{n_x}(A)^{n_x-k}B\Delta{\bf u}^{t+k-1}.
 \end{equation}
Denote ${\bf b}_m$ the $m$th column of the matrix $B$:
 \begin{equation}
 \label{eq_col_b}
 B=\left[\matrix{{\bf b}_1& {\bf b}_2& \cdots& {\bf b}_{n_u}}\right].
 \end{equation}
Regarding the terms $(A)^{n_x-k}{\bf b}_m$ as vectors in the $n_x$-dimensional
tangent space ${\mathcal T}$,
 \begin{equation}
 \label{eq_bas_set}
 {\bf h}^k_m=(A)^{n_x-k}{\bf b}_m,\quad k=1,\cdots,n_x,\ m=1,\cdots,n_u,
 \end{equation}
and the control perturbations $\Delta u_m^{t+k-1}$ as coordinates, we
immediately conclude that equation (\ref{eq_iter_gen}) rewritten as
 \begin{equation}
 \Delta{\bf x}_f-(A)^{n_x}\Delta{\bf x}_i=
 \sum_{k=1}^{n_x}\sum_{m=1}^{n_u}\Delta u_m^{t+k-1}{\bf h}^k_m
 \end{equation}
can only be satisfied, if and only if there are $n_x$ linearly independent
vectors in the set (\ref{eq_bas_set}), i.e., the set $\{{\bf h}^k_m\}$ spans
the tangent space ${\mathcal T}$. This is equivalent to requiring that
 \begin{equation}
 \label{eq_cont}
 {\rm rank}({\mathcal C})=n_x,
 \end{equation}
where the matrix
 \begin{equation}
 \label{eq_cont_mat}
  {\mathcal C}\equiv\left[\matrix{B& AB& (A)^2B & \cdots& (A)^{n_x-1}B}\right]
 \end{equation}
is called the {\em controllability matrix}. Condition (\ref{eq_cont}) was
introduced into the physics literature from linear systems theory by Romeiras
{\em et al.} \cite{romeiras} as a simple, but practical test of the
controllability. 

In contrast, the stabilizability condition requires that the set
(\ref{eq_bas_set}) spans only the unstable subspace $L^u\subseteq{\mathcal T}$
of the Jacobian $A$, instead of the whole tangent space ${\mathcal T}$.
Stabilizability can be formally expressed in the form identical to
(\ref{eq_cont}). Let us define the number of stable and unstable\footnote{For
the purpose of control we regard the central directions, defined by the
eigenvalues $\lambda$ such that $|\lambda|=1$ (${\rm Re}(\lambda)=0$ in the
continuous-time case), as unstable.} eigenvalues of the Jacobian $n_x^s$ and
$n_x^u$, respectively (one obviously has $n_x^s+n_x^u=n_x$). For instance, if
$A$ is a diagonalizable matrix, it has $n_x^s$ linearly independent stable
eigenvectors which we denote ${\bf e}^s_i$, $i=1,\cdots,n_x^s$. It can,
therefore, be shown using an appropriate coordinate transformation that the
pair $(A,B)$ is stabilizable if and only if
 \begin{equation}
 \label{eq_stab}
 {\rm rank}({\mathcal S})=n_x,
 \end{equation}
where the matrix
 \begin{equation}
 \label{eq_stab_mat}
  {\mathcal S}\equiv\left[\matrix{{\bf e}^s_1 & \cdots & {\bf e}^s_{n_x^s} &
                  B & AB & \cdots & (A)^{n_x^u-1}B}\right]
 \end{equation}
can be called the {\em stabilizability matrix} by analogy with the
controllability matrix.

In order to better understand the restrictions imposed on the control scheme by
symmetries, it is beneficial to look at the controllability condition from the
geometrical point of view, assuming $n_u=1$ and, consequently, $B={\bf b}$. 
The controllability in this context is equivalent to the vectors ${\bf h}^1,
{\bf h}^2,\cdots,{\bf h}^{n_x}$ spanning the tangent space ${\mathcal T}$.
Generically, the matrix $A$ is nondegenerate (has a nondegenerate spectrum), so
one can always find a vector ${\bf b}$, such that the resulting set
(\ref{eq_bas_set}) forms a basis. However, if $A$ is degenerate, which is a
usual consequence of symmetry, there will exist an eigenspace of the Jacobian,
$L^r\subset{\mathcal T}$, such that ${\bf x}^\dagger A=\lambda_r{\bf
x}^\dagger$, $\forall{\bf x}\in L^r$ with the dimension $d_r={\rm dim}(L^r)>1$,
where $\dagger$ denotes (complex conjugate) transpose of a matrix or vector.
The dynamics of the system in such an eigenspace cannot be controlled with just
one control parameter (see \cite{romeiras} for an example of such a situation),
because the vectors ${\bf h}^k$ only span a one-dimensional subspace of $L^r$.
Indeed, since $d_r>1$ there will exist $d_r-1$ adjoint eigenvectors ${\bf f}_j
\in L^r$ orthogonal to ${\bf b}$ and each other. Then
 \begin{equation}
 ({\bf f}_j\cdot{\bf h}^k)={\bf f}_j^\dagger(A)^{n_x-k}{\bf b}=
 \lambda_r^{n_x-k}{\bf f}_j^\dagger{\bf b}=
 \lambda_r^{n_x-k}({\bf f}_j\cdot{\bf b})=0,
 \end{equation}
so every basis vector ${\bf h}^k$ is orthogonal to every eigenvector ${\bf
f}_j$, $j=1,\cdots,d_r-1$.

It is often convenient to define the notion of controllability for individual
eigenvectors. We will say that the adjoint eigenvector ${\bf f}$ of the
Jacobian $A$ is controllable, if there exists $m$, $1\le m\le n_u$, such that
$({\bf f}\cdot{\bf b}_m)\ne 0$. Respectively, the eigenvector that is
orthogonal to every column of the control matrix $B$ is called uncontrollable.
Using these definitions we can, therefore, conclude that the controllability of
the linearized system is equivalent to the controllability of each and every
adjoint eigenvector of the Jacobian matrix (also see \cite{doyle}). Similarly,
the stabilizability is equivalent to the controllability of each and every {\em
unstable} adjoint eigenvector.

If the system dynamics in $L^r$ happens to be stable (e.g., when the system is
stabilizable, but uncontrollable), the system can still be stabilized similarly
to the nondegenerate case, but we have to ensure the controllability in case
the dynamics in this eigenspace is unstable. This can be achieved by increasing
the number of control parameters $n_u$, which extends the set
(\ref{eq_bas_set}), until it spans every eigenspace of ${\mathcal T}$. This
would lead one to assume that the minimal value of $n_u$ should be defined by
the highest degeneracy of the Jacobian matrix $A$. We will see, however, that
various kinds of degeneracy have a somewhat different effect on the
controllability of the system.

%%%%%%%%%%%%%%%%%%%%%%%%%%%%%%%%%%%%%%%%%%%%%%%%%%%%%%%%%%%%%%%%%%%%%%%%%%%%%%%
\subsection{Symmetries of the System}
\label{ss_symm}
%%%%%%%%%%%%%%%%%%%%%%%%%%%%%%%%%%%%%%%%%%%%%%%%%%%%%%%%%%%%%%%%%%%%%%%%%%%%%%%

Symmetries usually significantly simplify the analysis of system dynamics, and
the control problem is no exception. In particular, even when the exact form of
the Jacobian matrix is unknown, the structure of the symmetry group describing
the symmetries of the system allows one to reduce the controllability condition
(\ref{eq_cont}) to a set of much simpler conditions, which provide a number of
system-independent results. The discussion below is based on bifurcation theory
\cite{golubitsky} and closely parallels the treatment of degeneracy in quantum
mechanics and spontaneous symmetry breaking in quantum field theory and phase
transitions. 

In general we call the system symmetric, if the nonlinear evolution equation
preserves its form under a set of linear transformations $g:\,{\bf x}
\rightarrow{\bf x}'=g({\bf x})$ of the phase space. More formally, we say
that the evolution equation (\ref{eq_nl_id}) possesses a {\em structural}
symmetry described by a symmetry group ${\mathcal G}$, if the map ${\bf F}$
commutes with all group actions:
 \begin{equation}
 \label{eq_equiv}
 {\bf F}(g({\bf x}),{\bf u})=g({\bf F}({\bf x},{\bf u})),
 \quad\forall g\in {\mathcal G},\ \forall{\bf x}\in{\mathcal T}
 \end{equation}
or, in other words, if the function ${\bf F}({\bf x},{\bf u})$ is
${\mathcal G}$-equivariant with respect to its first argument. The group
${\mathcal G}$ is usually a byproduct of symmetries of the underlying physical
space, such as rotational and translational symmetry (domain symmetry), and
symmetries of the phase space, such as phase symmetry $\phi\rightarrow\phi+
2\pi$ (range symmetry). Since all interesting physical symmetries are unitary
(such rare exceptions as the Lorentz group are hardly relevant in the context
of control problem), we will assume that ${\mathcal G}$ is a unitary group.

Usually, the symmetry demonstrates itself in more than just one way: often
steady (as well as time-periodic) states $\bar{\bf x}$ of symmetric systems too
will be symmetric with respect to transformations $g\in{\mathcal H}_{\bar{\bf
x}}$, where ${\mathcal H}_{\bar{\bf x}}\subseteq{\mathcal G}$ is an {\em
isotropy} subgroup of $\bar{\bf x}$. In general, the target state $\bar{\bf x}$
might also be symmetric with respect to transformations which do not belong to
${\mathcal G}$. However, considering those does not provide any additional
information, so we assume that
 \begin{equation}
 \label{eq_invar}
 g(\bar{\bf x})=\bar{\bf x},\quad\forall g\in {\mathcal H}_{\bar{\bf x}}.
 \end{equation}
For the purpose of control it is important to observe that upon linearization
about the target state $\bar{\bf x}$ the structural symmetry of the evolution
equation (\ref{eq_nl_id}) does not disappear, but is replaced with a related
{\em dynamical} symmetry. Indeed, using the definitions (\ref{eq_equiv}),
(\ref{eq_invar}) and the fact that symmetry transformations are linear, one
obtains in the linear approximation for an arbitrary $g\in{\mathcal
H}_{\bar{\bf x}}$:
 \begin{eqnarray}
 \label{eq_chain}
 \bar{\bf x}+g(A\Delta{\bf x})
 &=&g(\bar{\bf x})+g(A\Delta{\bf x})
 =g(\bar{\bf x}+A\Delta{\bf x})\cr
 &=&g({\bf F}(\bar{\bf x},\bar{\bf u})+A\Delta{\bf x})
 =g({\bf F}(\bar{\bf x}+\Delta{\bf x},\bar{\bf u}))\cr
 &=&{\bf F}(g(\bar{\bf x}+\Delta{\bf x}),\bar{\bf u})
 ={\bf F}(g(\bar{\bf x})+g(\Delta{\bf x}),\bar{\bf u})\cr
 &=&{\bf F}(\bar{\bf x}+g(\Delta{\bf x}),\bar{\bf u})
 ={\bf F}(\bar{\bf x},\bar{\bf u})+Ag(\Delta{\bf x})\cr
 &=&\bar{\bf x}+Ag(\Delta{\bf x}).
 \end{eqnarray}
 
Defining ${\mathcal L}$ the full symmetry group of the linearized equation
(\ref{eq_lin_id_const}) in the absence of control ($\Delta{\bf u}^t={\bf 0}$):
 \begin{equation}
 \label{eq_symm}
 g(A\Delta{\bf x})=Ag(\Delta{\bf x}),\quad\forall g\in {\mathcal L},
 \end{equation}
one concludes that the group ${\mathcal L}$ describing the dynamical symmetry
of the system in the vicinity of the target state $\bar{\bf x}$ includes all
transformations $g\in{\mathcal H}_{\bar{\bf x}}$, and therefore:
 \begin{equation}
 \label{eq_g_prime}
 {\mathcal H}_{\bar{\bf x}}\subseteq{\mathcal L}.
 \end{equation}
One can speculate that typically ${\mathcal L}$ will coincide with ${\mathcal
H}_{\bar{\bf x}}$. As a consequence, if the target state $\bar{\bf x}$ has low
symmetry, the symmetry of the evolution equation will be reduced upon
linearization to a subgroup of ${\mathcal G}$. However, as we will see in
section \ref{s_examples}, ${\mathcal L}$ might be equal to ${\mathcal G}$,
or even include ${\mathcal G}$ as a subgroup for highly symmetric target
states, with the apparent symmetry increased by linearization.

It turns out that with the help of group representation theory one can
substantially simplify the controllability condition (\ref{eq_cont}) and, as a
result, obtain a number of useful restrictions on the set of control
parameters. Consider the matrix representation $T$ generated in the tangent
space ${\mathcal T}$ by the action of transformations $g$ from an
arbitrary subgroup ${\mathcal L}'$ of the full dynamical symmetry group
${\mathcal L}$:
 \begin{equation}
 \label{eq_in_rep}
 (g({\bf x}))_i=(T(g){\bf x})_i=\sum_{j=1}^{n_x}T_{ij}(g)x_j,\quad
 \forall{\bf x}\in {\mathcal T},
 \end{equation}
where, according to (\ref{eq_symm}), all matrices $T(g)$ commute with the
Jacobian
 \begin{equation}
 \label{eq_commut}
 T(g)A=AT(g),\quad \forall g\in {\mathcal L}'\subseteq{\mathcal L}.
 \end{equation}

The knowledge of the representation $T$ is enough to derive a very simple
criterion for the admissibility of the control matrix. Observe that, if
$T(g)B=B$ for an arbitrary transformation $g\in{\mathcal L}'$, then
 \begin{eqnarray}
 {\mathcal C}&=&\left[\matrix{T(g)g& AT(g)B& \cdots& (A)^{n_x-1}T(g)B}\right]
 \cr&=&\left[\matrix{T(g)B& T(g)AB& \cdots& T(g)(A)^{n_x-1}B}\right]
 =T(g){\mathcal C}.
 \end{eqnarray}
As a result, since $T_{ij}(g)\ne\delta_{i,j}$ for any $g\ne e$ (where we
defined $e$ as the identity transformation: $e({\bf x})={\bf x}$), the rows
$\tilde{\bf c}_j$ of the controllability matrix become linearly dependent,
 \begin{equation}
 \sum_{j=1}^{n_x}(T_{ij}(g)-\delta_{i,j})\tilde{\bf c}_j={\bf 0},
 \end{equation}
and the controllability condition (\ref{eq_cont}) is violated. Therefore, we
obtain a necessary condition on the control matrix:
 \begin{equation}
 \label{eq_invar_b}
 T(g)B\ne B,\quad\forall g\in{\mathcal L}'\setminus\{e\}.
 \end{equation}
In other words, the control arrangement should be chosen such that the symmetry
of the linearized evolution equation (\ref{eq_lin_id_const}) is completely
broken for (almost all) nonzero control perturbations $\Delta{\bf u}\ne{\bf
0}$.

%%%%%%%%%%%%%%%%%%%%%%%%%%%%%%%%%%%%%%%%%%%%%%%%%%%%%%%%%%%%%%%%%%%%%%%%%%%%%%%
\subsection{Group Coordinates}
\label{ss_diag}
%%%%%%%%%%%%%%%%%%%%%%%%%%%%%%%%%%%%%%%%%%%%%%%%%%%%%%%%%%%%%%%%%%%%%%%%%%%%%%%

Though simple and general, criterion (\ref{eq_invar_b}) is not very helpful for
finding the minimal set of control parameters satisfying the controllability
condition. In order to derive a more practically useful criterion one has to
make a few more steps. We begin with the reduction of the controllability
condition to a set of simpler conditions which can be performed \cite{rubio} by
constructing the Jordan block decomposition of the Jacobian matrix. This task
can be greatly simplified by transforming to the ``group coordinates,'' defined
with respect to the basis set composed of vectors which transform according to
different irreducible representations contained in $T$, in which the Jacobian
is block-diagonal. In practice, it is usually impossible to determine whether
the isotropy group ${\mathcal H}_{\bar{\bf x}}$ exhausts the dynamical
symmetries of the system or the group ${\mathcal L}$ contains some hidden
symmetries as well. It is, therefore, important to show that a number of
restrictions on the set of control parameters can be obtained using an
arbitrary unitary subgroup ${\mathcal L}'$ of ${\mathcal L}$.

Decomposing the representation $T$ into a sum of irreducible representations
$T^r$ of the group ${\mathcal L}'$ with respective dimensionalities $d_r$, we
obtain:
 \begin{equation}
 \label{eq_decomp}
 T=p_1T^1\oplus p_2T^2\oplus\cdots\oplus p_qT^q
 \end{equation}
with
 \begin{equation}
 \label{eq_dim_dec}
 n_x=p_1d_1+p_2d_2+\cdots+p_qd_q,
 \end{equation}
where $p_r$ denotes the number of equivalent representations $T^r$ present in
the decomposition (\ref{eq_decomp}), and $q$ is the total number of
nonequivalent irreducible representations. Since ${\mathcal L}'$ is unitary,
all irreducible representations $T^r$ in (\ref{eq_decomp}) can be chosen as
unitary \cite{hammer}.

The tangent space ${\mathcal T}$ is similarly decomposed into a sum of
invariant subspaces $L^{r\alpha}_{{\mathcal L}'}$ such that $T(g){\bf x}\in
L^{r\alpha}_{{\mathcal L}'}$, $\forall{\bf x}\in L^{r\alpha}_{{\mathcal L}'}$
and $\forall g\in {\mathcal L}'$:
 \begin{equation}
 \label{eq_inv_subsp}
 {\mathcal T}=L^1_{{\mathcal L}'}\oplus L^2_{{\mathcal L}'}
           \oplus\cdots\oplus L^q_{{\mathcal L}'},
 \end{equation}
where
 \begin{equation}
 \label{eq_inv_subsp_subsp}
 L^r_{{\mathcal L}'}=L^{r1}_{{\mathcal L}'}\oplus L^{r2}_{{\mathcal L}'}
           \oplus\cdots\oplus L^{rp_r}_{{\mathcal L}'}
 \end{equation}
and $\alpha=1,\cdots,p_r$ indexes different invariant subspaces, which
correspond to the same group of equivalent irreducible representations $T^r$.
It should be noted that even though the decomposition (\ref{eq_inv_subsp}) is
unique, the decomposition (\ref{eq_inv_subsp_subsp}) is not, unless $p_r=1$.
Let us introduce a basis in each invariant subspace $L^{r\alpha}_{{\mathcal
L}'}$ and denote the basis vectors ${\bf e}_i^{r\alpha}$, $i=1,\cdots,d_r$. We
choose the basis vectors such that they transform according to the irreducible
representation $T^r$, i.e.,
 \begin{equation}
 \label{eq_dummy_2}
 T(g)\,{\bf e}_i^{r\alpha}=\sum_{j=1}^{d_r}T^r_{ij}(g)\,{\bf e}_j^{r\alpha},
 \quad\forall g\in{\mathcal L}'.
 \end{equation}
For unitary $T^r$ a generalized orthogonality condition between basis vectors
${\bf e}_i^{r\alpha}$ can be established \cite{hammer} as a consequence of
(\ref{eq_dummy_2}):
 \begin{equation}
 ({\bf e}_i^{r\beta}\cdot{\bf e}_j^{s\alpha})=
 \delta_{r,s}\delta_{i,j}({\bf e}_i^{r\beta}\cdot{\bf e}_i^{r\alpha}).
 \end{equation}
In addition, for $p_r>1$ the decomposition (\ref{eq_inv_subsp_subsp}) can
always be performed in such a way that \hbox{$({\bf e}_i^{r\beta}\cdot{\bf
e}_i^{r\alpha})= \delta_{\alpha,\beta}$} (this, however, still leaves some
freedom in choosing the invariant subspaces $L^{r\alpha}_{{\mathcal L}'}$), so
that the complete set of basis vectors $\{{\bf e}_i^{r\alpha}\}$, where
$r=1,\cdots,q$, $\alpha=1,\cdots,p_r$, and $i=1,\cdots,d_r$ is made
orthonormal. We, therefore, conclude that the matrix $P$ defined by
 \begin{equation}
 \label{eq_p_mat}
 P=\left[\matrix{P^1\cr \vdots\cr P^q}\right],
         \qquad
 P^r=\left[\matrix{P^r_1\cr \vdots\cr P^r_{d_r}}\right],
         \qquad
 P^r_i=\left[\matrix{({\bf e}_i^{r1})^\dagger\cr 
           \vdots\cr ({\bf e}_i^{rp_r})^\dagger}\right],
 \end{equation}
is orthogonal, $(P)^{-1}=P^\dagger$ (or, more generally, unitary).
 
Furthermore, according to the Wigner-Eckart theorem \cite{hammer}, the matrix
elements of an arbitrary matrix (and the Jacobian $A$, in particular) invariant
with respect to any group transformation
 \begin{equation}
 T(g)AT^{-1}(g)=A,\quad\forall g\in{\mathcal L}',
 \end{equation}
satisfy the following general formula:
 \begin{equation}
 \label{eq_mat_elem}
 ({\bf e}_i^{r\beta}\cdot A{\bf e}_j^{s\alpha})=
 \delta_{r,s}\delta_{i,j}({\bf e}_i^{r\beta}\cdot A{\bf e}_i^{r\alpha}),
 \end{equation}
and the scalar product 
 \begin{equation}
 \label{eq_block_elem}
 (\bar{\Lambda}^r)_{\alpha\beta}\equiv({\bf e}_i^{r\alpha}\cdot
 A{\bf e}_i^{r\beta})
 \end{equation}
is independent of the index $i=1,\cdots,d_r$ (but depends on the decomposition
(\ref{eq_inv_subsp_subsp})). As a result, on transformation to the group
coordinates the Jacobian matrix becomes block diagonal:
 \begin{equation}
 \label{eq_jac_trans}
 \bar{A}=PAP^{-1}=\left[\matrix{
    \bar{A}^1 & & \cr
    & \ddots & \cr
    & &\bar{A}^q }\right],
 \end{equation}
where each block $\bar{A}^r$ is itself block-diagonal
 \begin{equation}
 \label{eq_blk_r}
 \bar{A}^r=\left[\matrix{
    \bar{\Lambda}^r & & \cr
    & \ddots & \cr
    & & \bar{\Lambda}^r }\right]
 \end{equation}
and consists of $d_r$ {\em identical} $p_r\times p_r$ blocks $\bar{\Lambda}^r$
with the matrix elements defined by the scalar product (\ref{eq_block_elem}).

If no irreducible representation $T^r$ of ${\mathcal L}'$ enters the
decomposition (\ref{eq_decomp}) more than once, i.e., $p_1=\cdots=p_q=1$, the
structure of the Jacobian matrix is completely resolved: the transformed
Jacobian is diagonal and its spectrum consists of eigenvalues $\lambda_r=
\bar{\Lambda}^r$, $r=1,\cdots,q$ with multiplicity $d_r$, while the basis
vectors ${\bf e}^{r\alpha}_i$ become the corresponding eigenvectors (and,
consequently, define the {\em normal modes} of the linearized system). In this
case the invariant subspaces of the group ${\mathcal L}'$ define the
eigenspaces of the Jacobian, $L^r=L^r_{{\mathcal L}'}$. Clearly, the spectrum
becomes degenerate, if the symmetry is sufficiently high (such that $T$
contains at least one irreducible representation $T^r$ with dimensionality
larger than one).

Degeneracy should not necessarily be associated with symmetry and might be
accidental (with respect to the group ${\mathcal L}'$). For instance, it can
happen that $\bar{\Lambda}^r=\bar{\Lambda}^{r'}$ for some $r\ne r'$, so that
the multiplicity of the eigenvalue $\lambda_r$ is increased respectively to
$d_r+d_{r'}$. Accidental degeneracies can be alternatively thought of as a
consequence of hidden symmetries contained in the full symmetry group
${\mathcal L}$ of which ${\mathcal L}'$ is a subgroup. However, the
degeneracies not associated with some physical symmetry are likely to disappear
under a typical perturbation, such as a change of system parameters and,
therefore, are most conveniently regarded as accidental. Since the full
symmetry group ${\mathcal L}$, in general, depends on system parameters and
cannot be directly deduced from the structural symmetry group ${\mathcal G}$,
it is usually more convenient to use its parameter-independent subgroup
${\mathcal L}'={\mathcal H}_{\bar{\bf x}}$ instead.

%%%%%%%%%%%%%%%%%%%%%%%%%%%%%%%%%%%%%%%%%%%%%%%%%%%%%%%%%%%%%%%%%%%%%%%%%%%%%%%
\subsection{Jordan Decomposition}
\label{ss_jordan}
%%%%%%%%%%%%%%%%%%%%%%%%%%%%%%%%%%%%%%%%%%%%%%%%%%%%%%%%%%%%%%%%%%%%%%%%%%%%%%%

If the symmetry described by ${\mathcal L}'$ is low, a number of equivalent
irreducible representations will typically be found in the decomposition
(\ref{eq_decomp}), i.e., we will have $p_r>1$ for certain $r$. In this case the
knowledge of the dynamical symmetries alone is not sufficient to completely
determine the structure of the Jacobian matrix, which is, in general,
system-dependent. As a result, one has to solve a secular equation
 \begin{equation}
 (\bar{\Lambda}^r-\lambda_{r\alpha}I){\bf e}^r_\alpha=0,
 \end{equation}
for each block $\bar{\Lambda}^r$ with $p_r>1$ in order to find the eigenvectors
in the invariant subspace $L^r_{{\mathcal L}'}$ and the respective eigenvalues.
Here, unlike the case of quantum mechanics, the Jacobian matrix does not have
to be Hermitian and, therefore, might not be diagonalizable. However,
$\bar{\Lambda}^r$ can always be reduced to the Jordan normal form by finding
the coordinate transformation $\bar{Q}^r$ such that
 \begin{equation}
 \label{eq_jord_dec}
 \Lambda^r=\bar{Q}^r\bar{\Lambda}^r(\bar{Q}^r)^{-1}=\left[\matrix{
    \Lambda^{r1} & & \cr
    & \ddots & \cr
    & & \Lambda^{rp'_r} }\right],
 \end{equation}
where $p'_r\le p_r$ is the number of distinct eigenvalues and the Jordan
superblock
 \begin{equation}
 \Lambda^{r\alpha}=\left[\matrix{
    \Lambda^{r\alpha}_1 & & \cr
    & \ddots & \cr
    & & \Lambda^{r\alpha}_{j_{r\alpha}} }\right]
 \end{equation}
corresponding to the eigenvalue $\lambda_{r\alpha}$ consists of $j_{r\alpha}$
Jordan blocks
 \begin{equation}
 \label{eq_jord_blk}
 \Lambda^{r\alpha}_i=\left[\matrix{
    \lambda_{r\alpha} & & & \cr
    1 & \lambda_{r\alpha} & & \cr
    & \ddots & \ddots & \cr
    & & 1 & \lambda_{r\alpha} }\right].
 \end{equation}
In the absence of accidental degeneracy all eigenvalues of $\bar{\Lambda}^r$
are different, so that $p'_r=p_r$ and $j_{r\alpha}=1$ for all $\alpha$, i.e.,
$\Lambda^r$ is diagonal.

Since each block $\bar{A}^r$ of the transformed Jacobian (\ref{eq_jac_trans})
consists of $d_r$ identical blocks $\bar{\Lambda}^r$, applying the coordinate
transformation defined by the block-diagonal matrix assembled from $d_r$ blocks
$\bar{Q}^r$,
 \begin{equation}
 \label{eq_tr_jordan}
 Q^r=\left[\matrix{
    \bar{Q}^r & & \cr
    & \ddots & \cr
    & & \bar{Q}^r }\right],
 \end{equation}
reduces $\bar{A}^r$ to the Jordan normal form:
 \begin{equation}
% \label{eq_jordan_blk}
 \tilde{A}^r=Q^r\bar{A}^r(Q^r)^{-1}=\left[\matrix{
    \Lambda^r & & \cr
    & \ddots & \cr
    & & \Lambda^r }\right].
 \end{equation}
The Jordan blocks on the diagonal of $\tilde{A}^r$ will not, in general, be
arranged in superblocks with the same eigenvalue. This, however, can be
trivially corrected by permuting the rows and columns of $\tilde{A}^r$ to
obtain the matrix
 \begin{equation}
 \label{eq_jordan_blk}
 \hat{A}^r=R^r\tilde{A}^r(R^r)^{-1}=\left[\matrix{
    \hat{A}^{r1} & & \cr
    & \ddots & \cr
    & & \hat{A}^{rp'_r} }\right],
 \end{equation}
where $R^r$ is the permutation matrix arranging the identical Jordan blocks
next to each other, and the Jordan superblock corresponding to the eigenvalue
$\lambda_{r\alpha}$ has the form
 \begin{equation}
 \label{eq_sup_blk}
 \hat{A}^{r\alpha}=\left[\matrix{
    \hat{A}^{r\alpha}_1 & & \cr
    & \ddots & \cr
    & & \hat{A}^{r\alpha}_{j_{r\alpha}} }\right].
 \end{equation}
Each block $\hat{A}^{r\alpha}_i$ is, in turn, composed of $d_r$ identical
Jordan blocks $\Lambda^{r\alpha}_i$, defined by (\ref{eq_jord_blk}):
 \begin{equation}
 \hat{A}^{r\alpha}_i=\left[\matrix{
    \Lambda^{r\alpha}_i & & \cr
    & \ddots & \cr
    & & \Lambda^{r\alpha}_i }\right].
 \end{equation}
Defining the block-diagonal coordinate transformation matrices $Q$ and $R$
 \begin{equation}
 \label{eq_trans_sr}
 Q=\left[\matrix{
    Q^1 & & \cr
    & \ddots & \cr
    & & Q^q }\right],\qquad
 R=\left[\matrix{
    R^1 & & \cr
    & \ddots & \cr
    & & R^q }\right],
 \end{equation}
we eventually obtain the sequence of coordinate transformations reducing the
Jacobian matrix $A$ to the Jordan normal form:
 \begin{equation}
 \label{eq_jord_pqr}
 \hat{A}=(RQP)A(RQP)^{-1}=\left[\matrix{
    \hat{A}^1 & & \cr
    & \ddots & \cr
    & & \hat{A}^q }\right],
 \end{equation}
where each block $\hat{A}^r$, $r=1,\cdots,q$ is defined by
(\ref{eq_jordan_blk}).

%%%%%%%%%%%%%%%%%%%%%%%%%%%%%%%%%%%%%%%%%%%%%%%%%%%%%%%%%%%%%%%%%%%%%%%%%%%%%%%
\subsection{Conditions for Controllability}
\label{ss_condition}
%%%%%%%%%%%%%%%%%%%%%%%%%%%%%%%%%%%%%%%%%%%%%%%%%%%%%%%%%%%%%%%%%%%%%%%%%%%%%%%

Once the Jacobian is reduced to the Jordan normal form, we can turn to the
problem of reducing the controllability condition to a set of simpler
conditions that will give us the restrictions on the admissible set of control
parameters. Since the controllability is a property of the system which does
not depend on the choice of the coordinate system, condition (\ref{eq_cont}) is
invariant with respect to any (nonsingular) coordinate transformation
\cite{rubio}, and hence is satisfied for the pair $(A,B)$, if and only if it is
satisfied for the pair $(\hat{A},\hat{B})$, where $\hat{B}=(RQP)B$ is the
transformed control matrix. Let us partition the transformed control matrix
$\hat{B}$ according to the block structure of $\hat{A}$:
 \begin{equation}
 \hat{B}=\left[\matrix{
 \hat{B}^1\cr\vdots\cr\hat{B}^q}\right],\qquad
 \hat{B}^r=\left[\matrix{
 \hat{B}^{r1}\cr\vdots\cr\hat{B}^{rp'_r}}\right],\qquad
 \hat{B}^{r\alpha}=\left[\matrix{\hat{B}^{r\alpha}_1\cr
 \vdots\cr\hat{B}^{r\alpha}_{j_{r\alpha}}}\right],\qquad
 \hat{B}^{r\alpha}_i=\left[\matrix{\hat{B}^{r\alpha}_{i1}\cr
 \vdots\cr\hat{B}^{r\alpha}_{id_r}}\right],
 \end{equation}
and denote $\hat{\bf b}^{r\alpha}_{ij}$ the first row of the matrix
$\hat{B}^{r\alpha}_{ij}$. Next, define the matrix $\bar{B}^{r\alpha}$ using the
relations
 \begin{equation}
 \label{eq_matr_bbar}
 \bar{B}^{r\alpha}=\left[\matrix{\bar{B}^{r\alpha}_1\cr
 \vdots\cr\bar{B}^{r\alpha}_{j_{r\alpha}}}\right],\qquad
 \bar{B}^{r\alpha}_i=\left[\matrix{\hat{\bf b}^{r\alpha}_{i1}\cr
 \vdots\cr\hat{\bf b}^{r\alpha}_{id_r}}\right].
 \end{equation}
In the absence of accidental degeneracy between the eigenvalues that correspond
to different invariant subspaces $L^r_{{\mathcal L}'}$ equation
(\ref{eq_sup_blk}) ensures that there are exactly $d_rj_{r\alpha}$ Jordan
blocks $\Lambda^{r\alpha}_i$ with the same eigenvalue $\lambda_{r\alpha}$. If,
however, there is such an accidental degeneracy involving $s$ different
invariant subspaces $L^{r_1}_{{\mathcal L}'},\cdots,L^{r_s}_{{\mathcal L}'}$,
such that for certain $\alpha_1,\cdots,\alpha_s$
 \begin{equation}
 \label{eq_eig_degen}
 \lambda_{r_1\alpha_1}=\cdots=\lambda_{r_s\alpha_s},
 \end{equation}
the number of Jordan blocks corresponding to the eigenvalue $\lambda_{r\alpha}$
increases to
 \begin{equation}
 j'_{r\alpha}\equiv\sum_{r',\alpha':\,\lambda_{r'\alpha'}=
 \lambda_{r\alpha}}d_{r'}j_{r'\alpha'}.
 \end{equation}
 
The knowledge of the number of Jordan blocks is very useful, since, according
to the standard result of linear system theory \cite{rubio}, it ultimately
determines the minimal number of control parameters. Specifically, it can be
shown that the controllability condition for the pair of matrices $(\hat{A},
\hat{B})$ is satisfied, if and only if for every $r$ and $\alpha$ (taken equal
to $r_1$ and $\alpha_1$ below)
 \begin{equation}
 \label{eq_row_vec_gen}
 {\rm rank}\,\left[\matrix{\bar{B}^{r_1\alpha_1}\cr\vdots
 \cr\bar{B}^{r_s\alpha_s}}\right]=j'_{r\alpha}=d_{r_1}j_{r_1\alpha_1}+
 \cdots+d_{r_s}j_{r_s\alpha_s},
 \end{equation}
where the indices $r_i$ and $\alpha_i$ are chosen according to
(\ref{eq_eig_degen}). This, in turn, can be achieved, if and only if $n_u\ge
j'_{r\alpha}$ for every $r$ and $\alpha$. Hence, in the most general case the
minimal number $\bar{n}_u$ of independent control parameters should equal the
maximal number of Jordan blocks with the same eigenvalue $\lambda_{r\alpha}$:
 \begin{equation}
 \label{eq_max_blk_gen}
 \bar{n}_u=\max_{r=1,\cdots,q}\max_{\alpha=1,\cdots,p'_r}j'_{r\alpha}.
 \end{equation}

Note that, since the block $\bar{B}^{r\alpha}$ has $d_rj_{r\alpha}$ rows, ${\rm
rank}(\bar{B}^{r\alpha})\le d_rj_{r\alpha}$ for every $r$ and $\alpha$. Using
this fact, the trivial matrix inequality
 \begin{equation}
 {\rm rank}\,\left[\matrix{\bar{B}^{r_1\alpha_1}\cr\vdots
 \cr\bar{B}^{r_s\alpha_s}}\right]\le {\rm rank}(\bar{B}^{r_1\alpha_1})+
 \cdots+{\rm rank}(\bar{B}^{r_s\alpha_s}),
 \end{equation}
and equation (\ref{eq_row_vec_gen}) one obtains
 \begin{equation}
 \label{eq_row_vec_red}
 {\rm rank}(\bar{B}^{r\alpha})=d_rj_{r\alpha},
 \quad r=1,\cdots,q,\ \alpha=1,\cdots,p_r.
 \end{equation}
Furthermore, according to the definition (\ref{eq_matr_bbar}) of the matrix
$\bar{B}^{r\alpha}$, for every $r$ and $\alpha$ ${\rm rank}(\hat{B}^{r\alpha})
\ge{\rm rank}(\bar{B}^{r\alpha})$, so one can write
 \begin{equation}
 \label{eq_dummy_1}
 {\rm rank}(\hat{B}^r)=
 {\rm rank}\,\left[\matrix{\hat{B}^{r1}\cr\vdots\cr\hat{B}^{rp_r}}\right]
 \ge\max_{\alpha=1,\cdots,p_r}{\rm rank}(\hat{B}^{r\alpha})
 \ge d_r\max_{\alpha=1,\cdots,p_r}j_{r\alpha}.
 \end{equation}
In addition, since $Q^r$ and $R^r$ are nonsingular coordinate transformations
which do not change the rank of a matrix,
 \begin{equation}
 \label{eq_rank_eq}
 {\rm rank}(\hat{B}^r)={\rm rank}(R^rQ^rP^rB)={\rm rank}(P^rB),
 \end{equation}
where we got rid of all system-specific information, which was contained in the
matrices $Q^r$ and $R^r$.

The symmetry information alone is insufficient to determine the values of
either $j_{r\alpha}$ or $j'_{r\alpha}$. However, by definition one has $p_r\ge
j_{r\alpha}\ge 1$ so that $j'_{r\alpha}\ge d_r$. As a consequence, we obtain
two necessary conditions for controllability. First of all, equation
(\ref{eq_max_blk_gen}) yields the lower bound on the minimal number of control
parameters
 \begin{equation}
 \label{eq_max_blk_ineq}
 \bar{n}_u\ge\max_{r=1,\cdots,q}d_r.
 \end{equation}
Second, inequality (\ref{eq_dummy_1}) combined with equality (\ref{eq_rank_eq})
imposes a number of restriction on the control matrix $B$,
 \begin{equation}
 \label{eq_row_vec_ineq}
 {\rm rank}(P^rB)\ge d_r,\quad r=1,\cdots,q,
 \end{equation}
which can be interpreted as the requirement of the mutual independence of
control parameters. We can therefore, conclude that an arbitrary (unitary)
subgroup ${\mathcal L}'$ of the full dynamical symmetry group ${\mathcal L}$
does not completely define the minimal set of control parameters. It does,
however, define a set of necessary conditions required for controllability. In
general, the knowledge of all dynamical symmetries, both unitary and
nonunitary, described by the group ${\mathcal L}$ is required in order to
completely resolve the structure of the Jacobian matrix and obtain the
necessary and sufficient condition for controllability.

Nevertheless, even without knowing the full symmetry group ${\mathcal L}$ one
can obtain the necessary and sufficient conditions by making a number of
assumptions. First, assume that there are no accidental degeneracies (it is
usually safe to do so if, e.g., ${\mathcal L}'$ is taken to coincide with
${\mathcal H}_{\bar{\bf x}}$: we ensure that all physical symmetries are taken
into account, and accidental degeneracies should only appear for certain
special values of system parameters). Then $j_{r\alpha}=1$, $j'_{r\alpha}=d_r$,
and $\bar{B}^{r\alpha}= \hat{B}^{r\alpha}$ for all $r$ and $\alpha$, so
condition (\ref{eq_max_blk_gen}) is equivalent to
 \begin{equation}
 \label{eq_max_blk}
 \bar{n}_u=\max_{r=1,\cdots,q}d_r.
 \end{equation}
If, in addition, no irreducible representation $T^r$ of ${\mathcal L}'$ enters
the decomposition (\ref{eq_decomp}) more than once, such that $p_r=1$ for all
$r$, instead of inequality (\ref{eq_row_vec_ineq}) one obtains the equality:
 \begin{equation}
 \label{eq_row_vec}
 {\rm rank}(P^rB)=d_r,\quad r=1,\cdots,q.
 \end{equation}

Conditions (\ref{eq_row_vec_ineq}) and (\ref{eq_row_vec}) can be simplified
even further by defining the projection operator \hbox{$\hat{P}^r\equiv
(P^r)^\dagger P^r$} onto the invariant subspace $L^r_{{\mathcal L}'}\subset
{\mathcal T}$. This operator can be obtained directly from the matrix
representation $T$ for most symmetry groups of interest. For finite discrete
groups it is given by
 \begin{equation}
 \label{eq_proj_disc}
 \hat{P}^r=\frac{d_r}{n_g}\sum_{g\in{\mathcal L}'}\chi^r(g) T(g),
 \end{equation}
where $n_g$ is the number of elements of the group ${\mathcal L}'$ and
$\chi^r(g)$ is the character of the group element $g$ in the representation
$T^r$. Similarly, for compact continuous groups we have
 \begin{equation}
 \label{eq_proj_cont}
 \hat{P}^r=d_r\int_{{\mathcal L}'}\chi^r(g) T(g)\,d\mu(g),
 \end{equation}
where $d\mu(g)$ is the group measure \cite{hammer}. Observing that ${\rm
rank}((P^r)^\dagger P^rB)={\rm rank}(P^rB)$, we can use the projection
operators to rewrite the condition (\ref{eq_row_vec}) in an equivalent form
 \begin{equation}
 \label{eq_rank_cnd}
 {\rm rank}(\hat{P}^rB)=d_r,\quad r=1,\cdots,q.
 \end{equation}

Summing up, we conclude that with the two assumptions made above the system is
controllable, if and only if the two conditions are met. The first one requires
the number $n_u$ of control parameters to be greater or equal to the
dimensionality $d_r$ of the largest irreducible representation $T^r$ present in
the decomposition of the matrix representation $T$ of the subgroup ${\mathcal
L}'\subseteq{\mathcal L}$ in the tangent space ${\mathcal T}$. The second
one requires the control parameters to be independent: the columns ${\bf b}_m$
of the control matrix $B$ have to be chosen such that $d_r$ of the projections
$\hat{P}^r{\bf b}_m$,  $m=1,\cdots,n_u$ are linearly independent (and,
therefore, span the eigenspace $L^r=L^r_{{\mathcal L}'}$) for every
$r=1,\cdots,q$. The last requirement imposes a number of restrictions on the
admissible form of the linear response of the system to perturbations of
control parameters.

A number of comments are in order. First of all, as we have just seen, the
number of control parameters is determined by the number of Jordan blocks with
the same eigenvalue, not the multiplicity of that eigenvalue. It becomes
intuitively clear why this is so, if one compares the action of different
Jacobians already reduced to the Jordan form. For instance, the Jacobian
 \begin{equation}
 A_1=\left[\matrix{
    \lambda & & \cr
    & \lambda & \cr
    & &\lambda }\right]
 \end{equation}
generates the set of three linearly dependent vectors ${\bf h}^0={\bf b}$,
${\bf h}^1=\lambda{\bf b}$, ${\bf h}^2=\lambda^2{\bf b}$ (compare to
(\ref{eq_bas_set})), that span a one-dimensional subspace of ${\mathbb R}^3$
for an arbitrary choice of ${\bf b}$. As a result, three control parameters and
a control matrix with three linearly independent columns, $B=\left[\matrix{{\bf
b}_1 & {\bf b}_2 & {\bf b}_3}\right]$, are necessary to control the system. On
the contrary, the Jacobian
 \begin{equation}
 A_2=\left[\matrix{
    \lambda & & \cr
    1 & \lambda & \cr
    & 1 &\lambda }\right]
 \end{equation}
generates a linearly independent set of basis vectors that spans ${\mathbb
R}^3$, requiring just one control parameter and a control matrix with a single
column $B={\bf b}$.

Second, symmetry does not always make the Jacobian degenerate, and the
nondegenerate case can be handled in the same way as the one with no
symmetries. Neither does the degeneracy by itself imply that multi-parameter
control is required: even if the eigenvalue $\lambda_{r'\alpha'}$ is
degenerate, but $j'_{r\alpha}=d_r=1$ for every $r$ and $\alpha$ (the degeneracy
is accidental and limited to a single invariant subspace $L^{r'}_{{\mathcal
L}'}$), one control parameter is sufficient to ensure the controllability. In
both cases, however, the dynamical symmetry should be rather low. Specifically,
the decomposition (\ref{eq_decomp}) of the matrix representation $T$ should not
contain any multi-dimensional irreducible representations.

Finally, the conditions on the set of control parameters that were obtained
above are imposed by the {\em controllability} condition and guarantee that
control can be achieved. However, in general, only the weaker {\em
stabilizability} condition has to be satisfied which, according to section
\ref{ss_stab_cont}, requires that every {\em unstable} normal mode of the
system is controllable, so that, only $r$ and $\alpha$ such that
$|\lambda_{r\alpha}|\le 1$ have to be considered in the conditions
(\ref{eq_row_vec_gen}) and (\ref{eq_max_blk_gen}). As a consequence, it might
be possible to stabilize highly symmetric states of compact extended systems
with strong spatial correlations using a single control parameter --- if only a
small number of modes is excited, there is a chance that all {\em unstable}
modes will correspond to one-dimensional irreducible representations $T^r$. In
strongly chaotic systems a large number of modes will be unstable and many of
them will inevitably be degenerate, calling for multi-parameter control.
Similar considerations apply to weakly chaotic systems with large spatial
extent.

%%%%%%%%%%%%%%%%%%%%%%%%%%%%%%%%%%%%%%%%%%%%%%%%%%%%%%%%%%%%%%%%%%%%%%%%%%%%%%%
\section{Time-Periodic States}
\label{s_ltv}
%%%%%%%%%%%%%%%%%%%%%%%%%%%%%%%%%%%%%%%%%%%%%%%%%%%%%%%%%%%%%%%%%%%%%%%%%%%%%%%

The results obtained above for the time-invariant case can be generalized for
the time-varying and, in particular, time-periodic case, but first we have to
define the notions of controllability and dynamical symmetry in the context of
time-varying trajectories. Indeed, in the time-varying case the Jacobian $A^t$
and the control matrix $B^t$ in the linearized evolution equation
(\ref{eq_lin_id}) are time-dependent and, as a consequence, neither the
definition of controllability given in section \ref{ss_stab_cont} nor the
condition (\ref{eq_cont}) holds. Besides, it is not at all clear that the
symmetry of the target trajectory, and hence the dynamical symmetry group
${\mathcal L}$ can be uniquely and consistently defined.

We will see that all these notions generalize in a rather straightforward way,
so that the same formalism as we used in the previous sections applies here as
well. To begin with, we define the controllability of a general time-varying
linear system. Expanding the definition given for time-invariant target states,
we call the $n_x$-dimensional linear system (\ref{eq_lin_id}) or the sequences
of matrices $\{A^t,B^t\}$ controllable if, for any initial state $\Delta{\bf
x}^{t_i}=\Delta{\bf x}_i$, times $t_f-t_i\ge n_x$, and final state $\Delta{\bf
x}_f$, there exists a sequence of control perturbations $\Delta{\bf
u}^{t_i},\cdots, \Delta{\bf u}^{t_f-1}$ such that the solution of equation
(\ref{eq_lin_id}) satisfies $\Delta{\bf x}^{t_f}=\Delta{\bf x}_f$.

The controllability condition can be restated in terms of the matrices $A^t$
and $B^t$ conducting the analysis similar to that of section
\ref{ss_stab_cont}. Applying the map (\ref{eq_lin_id}) $n_x$ times yields
 \begin{eqnarray}
 \label{eq_target}
 \Delta{\bf x}^{t+n_x}=J_{n_x}^{t+n_x-1}\Delta{\bf x}^t+\sum_{k=0}^{n_x-1}
 J_{n_x-1-k}^{t+n_x-1}B^{t+k}\Delta{\bf u}^{t+k},
 \end{eqnarray}
where we have introduced a shorthand notation
 \begin{equation}
 \label{eq_jacob_prod}
 J_k^t=A^tA^{t-1}\cdots A^{t-k+1}
 \end{equation}
for the product of $k$ consecutive Jacobians. Arguments identical to those used
to derive the controllability condition (\ref{eq_cont}) from equation
(\ref{eq_iter_gen}) allow us to conclude that for time-varying states the
controllability condition can again be written in the matrix form:
 \begin{equation}
 \label{eq_seqn_cont}
 {\rm rank}({\mathcal C}_t)=n_x,\quad \forall t,
 \end{equation}
where the controllability matrix (\ref{eq_cont_mat}) is now replaced with the
sequence of matrices
 \begin{equation}
 \label{eq_seqn_mat}
 {\mathcal C}_t\equiv\left[\matrix{B^t & J_1^tB^{t-1} & J_2^tB^{t-2} &
    \cdots & J_{{n_x}-1}^tB^{t-n_x+1}}\right].
 \end{equation}

Next we have to define the dynamic symmetry group ${\mathcal L}$. Suppose the
target trajectory $\bar{\bf x}^1,\bar{\bf x}^2,\cdots,\bar{\bf x}^\tau$ has
period $\tau$, and the symmetry of the point $\bar{\bf x}^t$ on the target
trajectory is described by the group ${\mathcal H}_{\bar{\bf x}^t}\subseteq
{\mathcal G}$. We can then write
 \begin{equation}
 g(\bar{\bf x}^{t+1})=g({\bf F}(\bar{\bf x}^t,\bar{\bf u}))=
 {\bf F}(g(\bar{\bf x}^t),\bar{\bf u})={\bf F}(\bar{\bf x}^t,\bar{\bf u})=
 \bar{\bf x}^{t+1}
 \end{equation}
 for every $g\in {\mathcal H}_{\bar{\bf x}^t}$. Consequently,
 \begin{equation}
 {\mathcal H}_{\bar{\bf x}^1}\subseteq {\mathcal H}_{\bar{\bf x}^2}
 \subseteq\cdots\subseteq{\mathcal H}_{\bar{\bf x}^\tau}
 \subseteq {\mathcal H}_{\bar{\bf x}^1},
 \end{equation}
which means that the symmetry properties of all the points on the target
trajectory are the same and the isotropy symmetry group of the {\em trajectory}
${\mathcal H}_{\bar{\bf x}}$ can be uniquely defined using an arbitrary point
$\bar{\bf x}^t$, ${\mathcal H}_{\bar{\bf x}}={\mathcal H}_{\bar{\bf x}^t}$.

Using the arguments that lead to equation (\ref{eq_chain}) we obtain for an
arbitrary $g\in{\mathcal H}_{\bar{\bf x}}$:
 \begin{eqnarray}
 \bar{\bf x}^{t+1}+g(A^t\Delta{\bf x})
 &=&g(\bar{\bf x}^{t+1})+g(A^t\Delta{\bf x})
 =g(\bar{\bf x}^{t+1}+A^t\Delta{\bf x})\cr
 &=&g({\bf F}(\bar{\bf x}^t,\bar{\bf u})+A^t\Delta{\bf x})
 =g({\bf F}(\bar{\bf x}^t+\Delta{\bf x},\bar{\bf u}))\cr
 &=&{\bf F}(g(\bar{\bf x}^t+\Delta{\bf x}),\bar{\bf u})
 ={\bf F}(g(\bar{\bf x}^t)+g(\Delta{\bf x}),\bar{\bf u})\cr
 &=&{\bf F}(\bar{\bf x}^t+g(\Delta{\bf x}),\bar{\bf u})
 ={\bf F}(\bar{\bf x}^t,\bar{\bf u})+A^tg(\Delta{\bf x})\cr
 &=&\bar{\bf x}^{t+1}+A^tg(\Delta{\bf x}).
 \end{eqnarray}
This, in turn, means that the symmetry group ${\mathcal L}_t$ of the Jacobian
$A^t$ satisfies
 \begin{equation}
 \label{eq_gt_prime}
 {\mathcal H}_{\bar{\bf x}}\subseteq{\mathcal L}_t,\quad t=1,\cdots,\tau.
 \end{equation}
Again, typically, we expect ${\mathcal L}_t={\mathcal H}_{\bar{\bf x}}$, so
that ${\mathcal L}$ too would be unique for any given periodic trajectory as
would the matrix representation $T$, such that
 \begin{equation}
 \label{eq_comm_tv}
 T(g)A^t=A^tT(g),\quad \forall g\in {\mathcal L}.
 \end{equation}
It is, therefore, enough to know the symmetry properties of an arbitrary point
of the periodic trajectory in order to establish the requirements on the
control scheme similarly to the time-invariant case. If ${\mathcal L}_t$ is not
unique, we can still use the commutation relation (\ref{eq_comm_tv}) for
the subgroup ${\mathcal L}'={\mathcal H}_{\bar{\bf x}}$ to obtain a lower bound
on the minimal number of control parameters.

Finally, we note that although it is possible to obtain certain results for
time-varying control matrices $B^t$, we assume, as is often the case in real
systems, that $B^t$ is constant and drop the time index. As we will discover
below, in the time-periodic case the restrictions imposed by symmetry on the
structure of the matrix $B$ can typically be determined without the detailed
knowledge of the Jacobian matrices, but based on the symmetry properties alone,
similarly to the time-invariant case. Indeed, let us construct the
representation $T$ of the group ${\mathcal L}'$ in the tangent space ${\mathcal
T}$ and decompose it into the sum of irreducible representations. This
again defines a set of invariant subspaces $L^{r\alpha}_{{\mathcal L}'}$ and a
set of basis vectors $\{{\bf e}^{r\alpha}_i\}$, which we use to construct the
coordinate transformation matrix $P$ according to the definition
(\ref{eq_p_mat}).

Since the rank of the matrix (\ref{eq_seqn_mat}) does not change under
a coordinate transformation, the controllability condition (\ref{eq_seqn_cont})
is equivalent to the condition
 \begin{equation}
 \label{eq_ctr_cnd_trn}
 {\rm rank}(\bar{\mathcal C}_t)=n_x,\quad t=1,\cdots,\tau
 \end{equation}
where
 \begin{equation}
 \bar{\mathcal C}_t=\left[\matrix{\bar{B} & \bar{J}_1^t\bar{B} &
    \cdots & \bar{J}_{{n_x}-1}^t\bar{B}}\right],
 \end{equation}
$\bar{B}=PB$ and $\bar{J}^t_k=PJ^t_k(P)^{-1}$. The products $J_k^t$ have the
same symmetry properties as the Jacobian matrices $A^t$ for arbitrary $k$ and
$t$, and, therefore, both the matrices $A^t$ and the products $J^t_k$
block-diagonalize in exactly the same way:
 \begin{equation}
 \bar{A}^t=PA^t(P)^{-1}=\left[\matrix{
    \bar{A}^{t,1} & & \cr
    & \ddots & \cr
    & &\bar{A}^{t,q} }\right],
 \end{equation}
and
 \begin{equation}
 \bar{J}^t_k=PJ^t_k(P)^{-1}=\left[\matrix{
    \bar{J}_k^{t,1} & & \cr
    & \ddots & \cr
    & &\bar{J}_k^{t,q} }\right].
 \end{equation}
Similarly to the time-invariant case, the blocks $\bar{A}^{t,r}$ and
$\bar{J}_k^{t,r}$ are themselves block-diagonal
 \begin{equation}
 \bar{A}^{t,r}=\left[\matrix{
    \bar{\Lambda}^{t,r} & & \cr
    & \ddots & \cr
    & & \bar{\Lambda}^{t,r} }\right],
 \qquad
 \bar{J}^{t,r}=\left[\matrix{
    \bar{\Gamma}^{t,r}_k & & \cr
    & \ddots & \cr
    & & \bar{\Gamma}^{t,r}_k }\right]
 \end{equation}
and consist of $d_r$ identical $p_r\times p_r$ blocks $\bar{\Lambda}^{t,r}$ and
$\bar{\Gamma}^{t,r}_k$, respectively, whose matrix elements are defined by the
scalar products
 \begin{eqnarray}
 (\bar{\Lambda}^{t,r})_{\alpha\beta}&\equiv&({\bf e}_i^{r\alpha}\cdot
 A^t{\bf e}_i^{r\beta}),\cr
 (\bar{\Gamma}^{t,r}_k)_{\alpha\beta}&\equiv&({\bf e}_i^{r\alpha}\cdot
 J^t_k{\bf e}_i^{r\beta}).
 \end{eqnarray}
Using the definition (\ref{eq_jacob_prod}) one can check that for any $t$, $k$
and $r$ the matrix $\bar{\Gamma}_k^{t,r}$ can be represented as the product
 \begin{equation}
 \bar{\Gamma}_k^{t,r}=\bar{\Lambda}^{t,r}\bar{\Lambda}^{t-1,r}\cdots
 \bar{\Lambda}^{t-k+1,r}.
 \end{equation}

Let us partition the transformed control matrix $\bar{B}$ into blocks
$\bar{B}^r=P^rB$ and define the reduced controllability matrices
 \begin{equation}
 \label{eq_red_cmtv}
 \bar{\mathcal C}^r_t=\left[\matrix{\bar{B}^r & \bar{J}_1^{t,r}\bar{B}^r &
    \cdots & \bar{J}_{n_x-1}^{t,r}\bar{B}^r}\right].
 \end{equation}
Using relation (\ref{eq_dim_dec}) and the fact that the matrix $\bar{\mathcal
C}^r_t$ has $d_rp_r$ rows one can write
 \begin{equation}
 {\rm rank}(\bar{\mathcal C}_t)={\rm rank}\,\left[\matrix{\bar{\mathcal C}^1_t
    \cr\vdots\cr\bar{\mathcal C}^q_t}\right]\le{\rm rank}(\bar{\mathcal
    C}^1_t)+\cdots{\rm rank}(\bar{\mathcal C}^q_t)\le d_1p_1\cdots d_qp_q=n_x
 \end{equation}
to obtain as a consequence of (\ref{eq_ctr_cnd_trn}) the set of reduced
controllability conditions
 \begin{equation}
 \label{eq_red_gcc}
 {\rm rank}(\bar{\mathcal C}_t^r)=d_rp_r,\quad r=1,\cdots,q.
 \end{equation}

The blocks $\bar{J}_k^{t,r}\bar{B}^r$ of the matrix (\ref{eq_red_cmtv}) can
become linearly dependent for certain $\tau$, $d_r$ and $p_r$. Indeed, it is
trivial to see that for a sequence of $n$ arbitrary $p\times p$ matrices $R_i$,
it is always possible to find a set of coefficients $\mu_0,\mu_1,\cdots,\mu_n$
such that
 \begin{equation}
 \label{eq_ann_gen}
 \mu_0I+\mu_1R_1+\cdots+\mu_nR_n=0,
 \end{equation}
as long as $n\ge p^2$. Equally easy to establish is the fact that, if the
matrices $R_i$ are not arbitrary, but satisfy the condition
 \begin{equation}
 \label{eq_sec_prod}
 R_i=W_1W_2\cdots W_i,
 \end{equation}
where $W_i$ is a sequence of arbitrary $p\times p$ matrices, such that
$W_{i+\tau}=W_i$, equation (\ref{eq_ann_gen}) can always be satisfied for
$n\ge\min(p^2,p\tau)$. The $p_r\times p_r$ matrices $\bar{\Gamma}_1^{t,r},
\cdots,\bar{\Gamma}_{n_x-1}^{t,r}$ form precisely the sequence satisfying the
condition (\ref{eq_sec_prod}). Besides, if the condition (\ref{eq_ann_gen}) is
satisfied for $R_i=\bar{\Gamma}_i^{t,r}$, it is satisfied for the sequence
$R_i=\bar{J}_i^{t,r}$ as well. As a result,
 \begin{equation}
 \label{eq_rnkc}
 {\rm rank}(\bar{\mathcal C}_t^r)={\rm rank}\left[\matrix{\bar{B}^r &
 \bar{J}_1^{t,r}\bar{B}^r & \cdots & \bar{J}_{n-1}^{t,r}\bar{B}^r}\right]
 \end{equation}
for $n=\min(p_r^2,p_r\tau)$ and arbitrary $\bar{B}^r$. Therefore, in order for
the conditions (\ref{eq_red_gcc}), and hence (\ref{eq_seqn_cont}), to be
satisfied, one should have
 \begin{equation}
 \label{eq_row_vec_tv}
 {\rm rank}(P^rB)\ge{\rm ceil}\left(\max\left(\frac{d_r}{p_r},
 \frac{d_r}{\tau}\right)\right),\quad r=1,\cdots,q,
 \end{equation}
where ${\rm ceil}(x)$ denotes the smallest integer number $n$ such that $n\ge
x$. The necessary conditions on the control matrix $B$, defined by
(\ref{eq_row_vec_tv}) are the generalization of the time-invariant result
(\ref{eq_row_vec_ineq}). Instead of (\ref{eq_max_blk_ineq}) one respectively
obtains the restriction on the minimal number of independent control parameters
required to satisfy the controllability condition (\ref{eq_seqn_cont}) for a
periodic target trajectory:
 \begin{equation}
 \label{eq_max_blk_tv}
 \bar{n}_u\ge{\rm ceil}\left(\max_{r=1,\cdots,q}\,\max
 \left(\frac{d_r}{p_r},\frac{d_r}{\tau}\right)\right).
 \end{equation}
It is interesting to note that a periodic trajectory can be made controllable
using the number of control parameters $n_u$ that could be smaller than the
number required for a steady state with the same symmetry.

Three special cases deserve separate consideration. First of all, suppose that
the Jacobian matrices $A^t$ commute with each other, so they can be
simultaneously diagonalized. In this case the condition (\ref{eq_ann_gen}) can
be satisfied by an appropriate choice of coefficients $\mu_1,\cdots,\mu_n$ for
$n\ge p_r$, so the necessary conditions (\ref{eq_max_blk_tv}) and
(\ref{eq_row_vec_tv}) will reduce to (\ref{eq_max_blk_ineq}) and
(\ref{eq_row_vec_ineq}), respectively, and $\bar{n}_u$ will no longer depend
on the period $\tau$ of the target trajectory.

Next, suppose there are no accidental degeneracies between the eigenvalues of
the Jacobians $A^t$ and their products $J^t_k$, and no irreducible
representation of ${\mathcal L}'$ appears in the decomposition
(\ref{eq_decomp}) more than once (so that Jacobian matrices can again be
simultaneously diagonalized). Now, however, identically to the time-invariant
case one obtains the necessary and sufficient conditions (\ref{eq_max_blk}) and
(\ref{eq_row_vec}) instead of the necessary conditions (\ref{eq_max_blk_ineq})
and (\ref{eq_row_vec_ineq}).

Finally, although we used the fact that the trajectory is periodic to derive
the above results, this requirement could be lifted, provided the symmetry of
all points on the target trajectory is the same, and, therefore, the condition
(\ref{eq_comm_tv}) is satisfied. A nonperiodic trajectory could then be treated
as a periodic one, with period $\tau=\infty$, and the condition
(\ref{eq_ann_gen}) will be satisfied by an appropriate choice of coefficients
$\mu_1,\cdots,\mu_n$ for $n\ge p_r^2$. As a result, instead of the restriction
(\ref{eq_max_blk_tv}) one will obtain
 \begin{equation}
 \bar{n}_u\ge{\rm ceil}\left(\max_{r=1,\cdots,q}\,\frac{d_r}{p_r}\right).
 \end{equation}

%%%%%%%%%%%%%%%%%%%%%%%%%%%%%%%%%%%%%%%%%%%%%%%%%%%%%%%%%%%%%%%%%%%%%%%%%%%%%%%
\section{Continuous-Time Systems}
\label{s_contin}
%%%%%%%%%%%%%%%%%%%%%%%%%%%%%%%%%%%%%%%%%%%%%%%%%%%%%%%%%%%%%%%%%%%%%%%%%%%%%%%

Most of the results obtained in the previous sections can be directly and
naturally generalized to continuous-time systems. This is a rather valuable
asset of the developed theory, since continuous-time control is, in general, a
much more flexible and powerful technique than discrete-time control. In the
presence of a decent continuous-time mathematical model (\ref{eq_nl_id_cont}),
continuous-time control can often achieve far superior results. It is, however,
a much more complicated technique as well. For simplicity we only discuss the
control of time-invariant target states. Linearizing the evolution equation
(\ref{eq_nl_id_cont}) around the steady target state $\bar{\bf s}$, one obtains
 \begin{equation}
 \label{eq_lin_cont}
 \Delta\dot{\bf s}(t)=A\Delta{\bf s}(t)+B\Delta{\bf u}(t),
 \end{equation}
where similarly to the discrete-time case we define the Jacobian
 \begin{equation}
 A={\bf D}_{\bf s}{\bf\Phi}(\bar{\bf s},\bar{\bf u})
 \end{equation}
and the control matrix 
 \begin{equation}
 B={\bf D}_{\bf u}{\bf\Phi}(\bar{\bf s},\bar{\bf u}).
 \end{equation}
The symmetries of the nonlinear evolution equation (\ref{eq_nl_id_cont}), the
target state $\bar{\bf s}$, and the linearization (\ref{eq_lin_cont}) are
determined identically to the discrete-time case using the relations
(\ref{eq_equiv}), (\ref{eq_invar}), and (\ref{eq_symm}), yielding the symmetry
groups ${\mathcal G}$, ${\mathcal H}_{\bar{\bf s}}$, and ${\mathcal L}$,
respectively. The definitions of the notions of stabilizability and
controllability in the continuous-time case are completely analogous to the
ones given in section \ref{ss_stab_cont} for the discrete-time case. 

The dynamical system described by equation (\ref{eq_lin_cont}) or the pair
$(A,B)$ is said to be controllable if, for any initial state $\Delta{\bf
s}(t_i)=\Delta{\bf s}_i$, times $t_f-t_i>0$ and final state $\Delta{\bf s}_f$,
there exists a (piecewise continuous) control perturbation $\Delta{\bf u}(t)$
such that the solution of equation (\ref{eq_lin_cont}) satisfies $\Delta{\bf
s}(t_f)=\Delta{\bf s}_f$. Otherwise the system or the pair $(A,B)$ is called
uncontrollable. 

Similarly, the dynamical system or the pair $(A,B)$ is said to be stabilizable,
if there exists a state feedback $\Delta{\bf u}(t)=-K\Delta{\bf s}(t)$ making
the system stable, such that all eigenvalues of the matrix $A'=A-BK$ have a
negative real part, ${\rm Re}(\lambda'_k)<0$, $\forall k$. Otherwise the system
or the pair $(A,B)$ is called unstabilizable.

The controllability of the pair $(A,B)$ again ensures that all eigenvalues of
$A'$ can be chosen appropriately, so that any controllable continuous-time
system is stabilizable as well. The controllability of a continuous-time system
is also established using the same criterion (\ref{eq_cont}) used to test for
the controllability in the discrete-time case. As a result, the conditions
imposed on the control matrix $B$ by the controllability condition in the
presence of symmetry are exactly the same as those obtained for discrete-time
systems.

%%%%%%%%%%%%%%%%%%%%%%%%%%%%%%%%%%%%%%%%%%%%%%%%%%%%%%%%%%%%%%%%%%%%%%%%%%%%%%%
\section{Symmetry Violation}
\label{s_violat}
%%%%%%%%%%%%%%%%%%%%%%%%%%%%%%%%%%%%%%%%%%%%%%%%%%%%%%%%%%%%%%%%%%%%%%%%%%%%%%%

In reality symmetries of physical systems displaying dynamical instabilities
are almost never exact. Indeed, the cylinders in a Taylor-Couette experiment
are never perfectly circular, the temperature inside a chemical reactor is
never absolutely uniform, neither are the rotor blades of a turbocompressor
exactly identical. The above analysis, on the other hand, has been conducted in
the assumption of exact symmetry. Therefore, it is essential to understand how
the obtained results change, if the symmetry is not exact or, in other words,
what the effect of a weak symmetry violation is. Such an analysis is also
crucial in the vicinity of points in the parameter space where symmetry
increasing bifurcations or accidental degeneracies occur.

For simplicity let us again consider the time-invariant case. The Jacobian $A$
of a weakly perturbed symmetric system takes the form
 \begin{equation}
 \label{eq_perturb_a}
 A=A_0+\epsilon A_1,
 \end{equation}
where $\epsilon$ denotes the magnitude of the perturbation and the unperturbed
Jacobian $A_0$ is exactly symmetric with respect to all transformations $g$ of
the group ${\mathcal L}$. For the group representation $T$ we thus have
 \begin{equation}
 T(g)A_0-A_0T(g)=0,\quad \forall g\in{\mathcal L}.
 \end{equation}
In general, the perturbation $\epsilon A_1$ will not be symmetric with respect
to any element of the group ${\mathcal L}$, except the identity transformation
$e$:
 \begin{equation}
 T(g)A_1-A_1T(g)\ne 0,\quad\forall g\in{\mathcal L}\setminus\{e\}.
 \end{equation}
Therefore, since
 \begin{equation}
 T(g)A-AT(g)=\epsilon(T(g)A_1-A_1T(g)),
 \end{equation}
the perturbation (\ref{eq_perturb_a}) completely destroys the symmetry of the
linearized evolution equation (\ref{eq_lin_id_const}) for any $\epsilon\ne 0$.
As a result, the perturbed system can be made controllable using a single
control parameter, irrespectively of the properties of the original symmetry
group ${\mathcal L}$. For instance, calculating the controllability matrix of
the perturbed system with $n_u=1$ and $B={\bf b}$ one obtains
 \begin{equation}
 {\mathcal C}={\mathcal C}_0+\epsilon {\mathcal C}_1+o(\epsilon^2),
 \end{equation}
where we defined
 \begin{eqnarray}
 {\mathcal C}_0&=&\left[\matrix{{\bf b}& A_0{\bf b}&\cdots& 
 (A_0)^{n_x-1}{\bf b}}\right],\cr
 {\mathcal C}_1&=&\left[\matrix{0 & A_1{\bf b}& \cdots& 
 ((A_0)^{n_x-2}A_1+\cdots+A_1(A_0)^{n_x-2}){\bf b}}\right].
 \end{eqnarray}
${\mathcal C}_0$ is clearly the controllability matrix of the unperturbed
system with full symmetry, which does not have a full rank, if the
decomposition (\ref{eq_decomp}) contains at least one irreducible
representation $T^r$ with the dimensionality $d_r>1$. Indeed, in the absence of
accidental degeneracies that would mean
 \begin{equation}
 \label{eq_unpt_rnk}
 n_0\equiv{\rm rank}({\mathcal C}_0)\le\sum_{r=1}^q p_r<n_x.
 \end{equation}
The controllability matrix ${\mathcal C}$ of the perturbed system, on the
other hand, has full rank for any $\epsilon\ne 0$ because the symmetry is
completely destroyed by the perturbation. Therefore, the perturbed {\em linear}
system becomes controllable even though the unperturbed system is not, for {\em
arbitrarily small} perturbations.

The controllability ensures that for any initial and final states of the linear
system (\ref{eq_lin_id_const}) the control can be found mapping the initial
state to the final state in $n_x$ iterations. Using (\ref{eq_iter_gen}) one
obtains explicitly
 \begin{equation}
 \label{eq_map_soln}
 \Delta{\bf U}^t\equiv\left[\matrix{\Delta u^{t+n_x-1}\cr\vdots\cr\Delta u^t}
 \right]=({\mathcal C})^{-1}(\Delta{\bf x}^{t+n_x}-(A)^{n_x}\Delta{\bf x}^t).
 \end{equation}
Formally, if the system is controllable, the controllability matrix is
invertible, and the solution (\ref{eq_map_soln}) is well defined for any
$\Delta{\bf x}^t$ and $\Delta{\bf x}^{t+n_x}$. However, when the matrix
${\mathcal C}$ is close to being singular its inverse is not well defined. It
is convenient to use the singular value decomposition of the controllability
matrix
 \begin{equation}
 {\mathcal C}=Q\Sigma R^\dagger,
 \end{equation}
where $Q=\left[\matrix{{\bf q}_1& {\bf q}_2& \cdots& {\bf q}_{n_x}}\right]$ and
$R=\left[\matrix{{\bf r}_1& {\bf r}_2& \cdots& {\bf r}_{n_x}}\right]$ are some
orthogonal $n_x\times n_x$ matrices, and
 \begin{equation}
 \Sigma=\left[\matrix{\sigma_1(\epsilon) & & \cr
                 & \ddots & \cr
                 & & \sigma_{n_x}(\epsilon)}\right].
 \end{equation}
The singular values are ordered such that $\sigma_1(\epsilon)\ge
\sigma_2(\epsilon)\ge\cdots\ge\sigma_{n_x}(\epsilon)$ for $\forall\epsilon$.
Additionally, equation (\ref{eq_unpt_rnk}) requires
 \begin{equation}
 \lim_{\epsilon\rightarrow 0}\sigma_i(\epsilon)=0,\quad i=n_0+1,\cdots,n_x.
 \end{equation}
In terms of the matrices $Q$, $\Sigma$, and $R$ we can write the inverse of
${\mathcal C}$ as
 \begin{equation}
 ({\mathcal C})^{-1}=R(\Sigma)^{-1}Q^\dagger=
 \sum_{i=1}^{n_x}\sigma_i^{-1}(\epsilon){\bf r}_i{\bf q}_i^\dagger
 \end{equation}
and, therefore, for small $\epsilon$ equation (\ref{eq_map_soln}) gives
 \begin{equation}
 \Delta{\bf U}^t\approx\sum_{i=n_0+1}^{n_x}
 \frac{({\bf q}_i\cdot\Delta{\bf x}^{t+n_x})
 -({\bf q}_i\cdot(A)^{n_x}\Delta{\bf x}^t)}
 {\sigma_i(\epsilon)}{\bf r}_i.
 \end{equation}
As a consequence, we obtain 
 \begin{equation}
 \lim_{\epsilon\rightarrow 0}|\Delta{\bf U}^t|=\infty.
 \end{equation}
This relation means that at least one control perturbation of the feedback
sequence $\Delta u^t,\cdots,\Delta u^{t+n_x-1}$ diverges as the symmetry
breaking perturbation $\epsilon A_1$ of the Jacobian vanishes. Since no
specific relation between the initial and the final state of the system was
implied, the obtained result is general, and does not depend on the control
method used to calculate the feedback. 

In fact, a more general statement holds. Suppose the symmetry is violated only
partially, such that the perturbed Jacobian (\ref{eq_perturb_a}) remains
exactly symmetric with respect to a subgroup ${\mathcal L}'$ of the full
symmetry group ${\mathcal L}$. Denote $\bar{n}_u$ and $\bar{n}'_u$ the minimal
number of control parameters required (assuming exact symmetry) by the groups
${\mathcal L}$ and ${\mathcal L}'$, respectively. Then it can be shown that,
similarly to the single-parameter case, at least one control perturbation of
the feedback sequence $\Delta{\bf u}^t,\cdots,\Delta{\bf u}^{t+n_x-1}$ diverges
as the symmetry breaking perturbation $\epsilon A_1$ of the Jacobian vanishes
whenever \hbox{$\bar{n}'_u\le n_u<\bar{n}_u$}. The same result is obtained if
the independent with respect to the group ${\mathcal L}'$ control parameters
become dependent with respect to the group ${\mathcal L}$, as indicated by the
violation of the general independence condition (\ref{eq_row_vec_gen}). The
time-periodic generalization is also straightforward. We will call this
situation {\em parametric deficiency}.

In other words, although it might be possible to control a {\em linear} system
with approximate symmetry using a number of control parameters which is smaller
than that required in the assumption of exact symmetry, the stabilization
requires feedback of very large magnitude. Such systems are called {\em weakly
controllable} in the language of control theory. However, the linear system is
only an abstraction. The linear approximation (\ref{eq_lin_id}) of the
evolution equation (\ref{eq_nl_id}) is only valid for small perturbations
$\Delta{\bf u}^t$ of the control parameters and small deviations $\Delta{\bf
x}^t$ from the target trajectory. Besides, additional restrictions on the
magnitude of the feedback are usually imposed by practical limitations, size
and energy constraints, etc., at the implementation stage. One can, therefore,
conclude that, since the feedback scales linearly with the deviation from the
target trajectory, a nonlinear system with parametric deficiency can be
stabilized using linear control only in an asymptotically contracting
neighborhood of the target trajectory.

Finally, consider the vicinity of the point $\bar{\bf u}_0$ in the parameter
space ${\mathbb R}^{n_u}$ at which an accidental degeneracy occurs, such that
the dynamical symmetry is described by the group ${\mathcal L}'$ for $\bar{\bf
u}\ne\bar{\bf u}_0$ and is increased to ${\mathcal L}$ (of which ${\mathcal
L}'$ is a subgroup) for $\bar{\bf u}=\bar{\bf u}_0$. In this case ${\mathcal
L}$ can be considered approximate symmetry in the vicinity of $\bar{\bf u}_0$,
and the distance to that point determines how strongly (or weakly) the symmetry
${\mathcal L}$ is violated. Suppose the control scheme is such that there is a
parametric deficiency. Then the system will remain controllable for $\bar{\bf
u}\ne \bar{\bf u}_0$. However, the strength of feedback required to control the
system will diverge as $\bar{\bf u}$ approaches $\bar{\bf u}_0$, at which point
the system will become uncontrollable.

%%%%%%%%%%%%%%%%%%%%%%%%%%%%%%%%%%%%%%%%%%%%%%%%%%%%%%%%%%%%%%%%%%%%%%%%%%%%%%%
\section{System Identification}
\label{s_recons}
%%%%%%%%%%%%%%%%%%%%%%%%%%%%%%%%%%%%%%%%%%%%%%%%%%%%%%%%%%%%%%%%%%%%%%%%%%%%%%%

Finally, let us consider the problem of phase space reconstruction (sometimes
also called system identification) in the context of symmetric systems. The
need to use phase space reconstruction stems from the fact that few
experimental systems exist whose dynamical equations are known (let alone with
precision required by the control algorithms). Even the complete information
about the state of these systems is rarely available, so one typically has to
contend with having a measurement of a single scalar output (sometimes called
an {\em observable}) for the description of the dynamics. This output, which we
call $y(t)$ is, in general, a function of the (unknown) internal state ${\bf
s}(t)$ of the system described by the (also unknown) continuous-time evolution
equation (\ref{eq_nl_id_cont}). In other words,
 \begin{equation}
 \label{eq_out_nl_id}
 y(t)=G({\bf s}(t)).
 \end{equation}

It turns out that it is possible to reconstruct both the internal state of the
system {\em and} its dynamics based on the time series measurement of the
output $y(t)$ using the procedure originally proposed by Packard {\em et al.}
\cite{packard}. The easiest way to recreate the state of the system from the
scalar output signal is to use time delays. Let us choose different delay times
$T_1,T_2,\cdots, T_{n_x}$ and construct an $n_x$-dimensional delay coordinate
vector
 \begin{equation}
 \label{eq_del_vec}
 {\bf x}(t)=\left[\matrix{y(t+T_1)\cr y(t+T_2)\cr\vdots\cr
 y(t+T_{n_x})}\right].
 \end{equation}
Takens showed \cite{takens} that for a scalar output (\ref{eq_out_nl_id}) and
conveniently chosen delay times $T_i$, if the dimension $n_x$ of the embedding
space is such that $n_x\ge 2n^h_s+1$, where $n^h_s$ is the Hausdorff dimension
of the attractor ${\mathcal A}$, the map ${\bf P}:\,{\bf s}(t)\rightarrow{\bf
x}(t)$ generically provides a global one-to-one representation of the attractor
and, hence, the system state. Finally, one recovers the map (\ref{eq_nl_id}) by
defining ${\bf x}^k={\bf x}(t_k)$, where $t_0, t_1, t_2, \cdots$ denote the
times of consecutive crossings of a conveniently chosen Poincar\'{e} surface of
section in the embedding space ${\mathbb R}^{n_x}$ by the reconstructed
trajectory ${\bf x}(t)$.

However, although the Takens' embedding theorem holds for a typical system
without symmetries, it is not satisfied for most of the symmetric systems due
to the symmetry-related degeneracy of the evolution operators. In other words,
most symmetric systems are {\em nongeneric} in the sense of Takens. As a result
of this nongenericity the state of the system becomes impossible to reconstruct
using a single scalar output, no matter how large the dimensionality $n_x$ of
the embedding space is, locally in the vicinity of highly symmetric periodic
trajectories. The attractor of the system remains folded at certain points and
along certain curves in the phase space, which prevents the global
reconstruction as well. Using the language of control theory we will say that
such systems are {\em unobservable} locally as well as globally.

The question of symmetry-caused nongenericity in the framework of phase space
reconstruction of a general symmetric system was first considered by King and
Stewart \cite{king}, who determined that the reason for the failure of the
embedding theorem is the violation of one of Takens' assumptions that the flow
defined by equation (\ref{eq_nl_id_cont}) has simple eigenvalues for low-period
periodic trajectories. As we have seen above, symmetric systems typically (but
not always) have degenerate eigenvalues (due to the fact that most nontrivial
irreducible representations are multi-dimensional) and, as a consequence, are
nongeneric. King and Stewart went on to formulate and prove a generalization of
the Takens' embedding theorem, which required the output to be a {\em vector},
not a {\em scalar}, function of the actual state of the system ${\bf s}(t)$:
 \begin{equation}
 \label{eq_vec_out_nl_id}
 {\bf y}(t)={\bf G}({\bf s}(t)),
 \end{equation}
mapping the phase space ${\mathcal Q}$ of the original system onto an
$n_y$-dimensional Euclidean space. The state of the system can then be
similarly represented by a delay coordinate vector
 \begin{equation}
 \label{eq_vec_del_vec}
 {\bf x}(t)=\left[\matrix{{\bf y}(t+T_1)\cr {\bf y}(t+T_2)\cr\vdots\cr 
 {\bf y}(t+T_{n_e})}\right],
 \end{equation}
where now the dimensionality of the embedding space is $n_x=n_yn_e$. The
question we have to answer is what conditions should the function ${\bf G}$
satisfy in order to allow a local (or global) one-to-one embedding. Since
the exact form of the evolution equations is assumed unknown, in order to find
the answer one can only exploit the symmetry properties of the system, which
are often easy to establish based on the underlying symmetries of the physical
space. Fortunately, the symmetry provides most of the necessary information.

Since we are interested in the issue of phase space reconstruction only as far
as it applies to the problem of linear control, we will assume a local
character for the observability property, unless explicitly stated otherwise.
According to the analysis conducted in \cite{king}, local embedding in the
vicinity of the periodic trajectory $\bar{\bf s}(t)$ requires ${\mathbb
R}^{n_y}$ to contain at least one copy of every invariant subspace
$L^{r\alpha}_{{\mathcal L}'}$ generated by the (unitary) irreducible
representation $T^r$ of the respective isotropy symmetry group ${\mathcal
L}'={\mathcal H}_{\bar{\bf s}}$. This would lead one to assume that the minimal
dimension $n_y$ of the output signal should be determined by the dimension of
the largest irreducible representation $T^r$.

This assumption can be trivially verified using the formalism developed above
for the control problem in the presence of symmetry. Indeed, let us again
consider a time-invariant target state $\bar{\bf s}$. Linearizing the output
(\ref{eq_vec_out_nl_id}) in the vicinity of this target state and denoting the
displacement $\Delta{\bf y}(t)={\bf G}({\bf s}(t))-{\bf G}(\bar{\bf s})$ one
obtains:
 \begin{equation}
 \label{eq_out_lin_cont}
 \Delta{\bf y}(t)=C\Delta{\bf s}(t),
 \end{equation}
where the constant matrix $C$ is defined thus:
 \begin{equation}
 C={\bf D}_{\bf s}{\bf G}(\bar{\bf s}).
 \end{equation}
The dynamical system defined by equations (\ref{eq_lin_cont}) and
(\ref{eq_out_lin_cont}) or the pair $(A,C)$ is said to be {\em observable} if,
for any times $t_f-t_i>0$, the initial state $\Delta{\bf s}(t_i)=\Delta{\bf
s}_i$ can be determined from the measurement of control perturbation
$\Delta{\bf u}(t)$ and output $\Delta{\bf y}(t)$ in the interval
$t\in[t_i,t_f]$. Otherwise, the system or the pair $(A,C)$ is said to be
unobservable.

It can be shown that the notion of observability is dual to the notion of
controllability. The crucial benefit of this duality is the fact that the
observability condition for the pair $(A,C)$ is equivalent \cite{rubio} to the
controllability condition for the pair $(A^\dagger,C^\dagger)$. Since the
commutation relation (\ref{eq_commut}) directly implies that
 \begin{equation}
 T(g)^\dagger A^\dagger=A^\dagger T(g)^\dagger,\quad\forall g\in{\mathcal L}',
 \end{equation}
the symmetry properties of the matrices $A$ and $A^\dagger$ are essentially
identical (as are the structures of their spectra, Jordan normal forms, etc.).
As a result, all restriction imposed on the control matrix $B$ by the
controllability condition in the presence of symmetry should be satisfied for
the matrix $C^\dagger$ as well. 

For instance, in case there are no accidental degeneracies and the
representation $T$ contains at most one copy of each irreducible representation
of the group ${\mathcal L}'$, one can claim that in order to reconstruct the
dynamics in the vicinity of the time-invariant symmetric target state $\bar{\bf
s}$ the number $n_y$ of measured scalar output signals $y_i(t)$ should be the
same as the minimal number $\bar{n}_u$ of independent control parameters, i.e.,
$\bar{n}_y=\bar{n}_u$. Furthermore, the outputs have to be independent, so that
$d_r$ of the projections $\hat{P}^r{\bf c}_i$, $i=1,\cdots,n_y$ are linearly
independent for every $r$, where $\hat{P}^r$ is the projection operator defined
by (\ref{eq_proj_disc}) and (\ref{eq_proj_cont}), and ${\bf c}_i^\dagger$ is
the $i$th row of the matrix $C$, which imposes a number of restrictions on the
allowed form of the function $G$. Local observability can be defined in the
same way for discrete-time systems. Careful consideration shows that symmetry
produces similar constraints independent of the particular description.

In addition to the notion of controllability of individual eigenvectors it is
often also convenient to define the notion of their observability. We will say
that the eigenvector ${\bf e}$ of the Jacobian $A$ is observable, if there
exists $i$, $1\le i\le n_y$, such that $({\bf e}\cdot{\bf c}_i)\ne 0$.
Respectively, an eigenvector that is orthogonal to every row of the matrix $C$
is called unobservable. Clearly, the observability of the linearized system is
equivalent to the observability of each and every eigenvector of the Jacobian
matrix.

In the conclusion of this section we make a few comments regarding the problem
of global phase space reconstruction. Often it is important to know how the
symmetry of the continuous-time experimental system transpires in the structure
of the discrete-time map (\ref{eq_nl_id}) obtained as a result of the time
delay embedding produced by a general output signal (\ref{eq_vec_out_nl_id}).
King and Stewart \cite{king} recognized that it is as important to preserve the
symmetry of the attractor as it is to preserve its topology during the
reconstruction. According to (\ref{eq_vec_out_nl_id}), using an arbitrary
vector output ${\bf y}(t)$ to generate the delay coordinate representation of
the system state corresponds to picking a function ${\bf G}$ which, in general,
distorts the symmetry. In order to preserve the symmetry of the original
attractor the function ${\bf G}$ has to be ${\mathcal G}$-equivariant:
 \begin{equation}
 {\bf G}(g({\bf s}))=g({\bf G}({\bf s})),\quad\forall g\in{\mathcal G},
 \ \forall{\bf s}\in{\mathcal Q},
 \end{equation}
where ${\mathcal G}$ is the structural symmetry group of the system
(\ref{eq_nl_id_cont}) which will, in general, act differently in the phase
space ${\mathcal Q}$ and Euclidean space ${\mathbb R}^{n_y}$. In addition, the
dimensionality $n_y$ of the Euclidean space has to be chosen high enough to
avoid local folding (obviously, $n_y$ should be no smaller than the number
$\bar{n}_y$ evaluated for the steady or periodic trajectory with the highest
isotropy symmetry). Finally, a global one-to-one embedding can be achieved by
choosing $n_e\ge 2n_s^h+1$ to preserve the topology of the attractor. The map
(\ref{eq_nl_id}) constructed using this embedding will preserve all dynamical
symmetries of the original system. However, the structural symmetry of the
differential equation (\ref{eq_nl_id_cont}) and the map (\ref{eq_nl_id}) will,
in general, be different.

%%%%%%%%%%%%%%%%%%%%%%%%%%%%%%%%%%%%%%%%%%%%%%%%%%%%%%%%%%%%%%%%%%%%%%%%%%%%%%%
\section{Examples}
\label{s_examples}
%%%%%%%%%%%%%%%%%%%%%%%%%%%%%%%%%%%%%%%%%%%%%%%%%%%%%%%%%%%%%%%%%%%%%%%%%%%%%%%

%%%%%%%%%%%%%%%%%%%%%%%%%%%%%%%%%%%%%%%%%%%%%%%%%%%%%%%%%%%%%%%%%%%%%%%%%%%%%%%
\subsection{Coupled Map Lattice}
\label{ss_cml}
%%%%%%%%%%%%%%%%%%%%%%%%%%%%%%%%%%%%%%%%%%%%%%%%%%%%%%%%%%%%%%%%%%%%%%%%%%%%%%%

Now that the formal theory is constructed, we can illustrate it by applying to
a few simple symmetric systems. First, consider the deterministic coupled map
lattice (CML) with nearest neighbor diffusive coupling \cite{kaneko}, which is
described by the following evolution equation
 \begin{equation}
 \label{eq_cml}
 x_i^{t+1}=\epsilon f(x_{i-1}^t,a)+(1-2\epsilon)f(x_i^t,a)
 +\epsilon f(x_{i+1}^t,a).
 \end{equation}
Coupled map lattices of this type are often used as models for such
spatiotemporally chaotic phenomena as surface growth, population dynamics, and
turbulence.  Here the index $i=1,2,\cdots,n_x$ labels the lattice sites, and
the periodic boundary condition is imposed. The choice of the map function
$f(x,a)$ is usually motivated by the local dynamics of the physical system
under consideration. In principle $f(x,a)$ can be chosen as an arbitrary
(nonlinear) function with parameter $a$, which typically represents the process
of generation and growth of local fluctuations, while diffusive coupling
typically plays the opposite role of dissipating these fluctuations. Therefore,
the parameters $a$ and $\epsilon$ specify the degree of instability and the
strength of dissipation in the system, respectively. For the purpose of
control, however, details of the local map are not important. The only aspect
of the control problem affected by any particular choice is the set of existing
unstable periodic trajectories.

The analysis of the controllability condition conducted in sections
\ref{s_steady} and \ref{s_ltv} shows that, if the system is symmetric, certain
symmetry-imposed restrictions on the choice of control parameters should be
satisfied in order to achieve control. In fact, the coupled map lattice is by
construction highly symmetric. The symmetry is that of the spatial lattice: the
evolution equation (\ref{eq_cml}) is invariant with respect to translations by
an integer number of lattice sites (periodic boundary condition makes the group
finite) and reflections about any site (or midplane between any adjacent
sites), which map the lattice back onto itself without destroying the adjacency
relationship. The corresponding point group ${\rm C}_{n_xv}$ (we assume $n_x$
-- even) has a total of $n_x/2+3$ nonequivalent irreducible representations.
The first four are one-dimensional, $d_1=d_2=d_3=d_4=1$, while the rest
$n_x/2-1$ are two-dimensional, $d_r=2$, $r\ge 5$. In comparison, breaking the
reflection symmetry reduces the group to ${\rm C}_{n_x}$, which only has
one-dimensional irreducible representations.

The dynamical symmetry group can be trivially obtained using the observation
that the Jacobian matrix in the linearization (\ref{eq_lin_id}) constructed
for the CML (\ref{eq_cml}) can always be represented as a product of two
matrices, $A^t=MN^t$, where
 \begin{equation}
 M_{ij}=(1-2\epsilon)\delta_{i,j}+\epsilon(\delta_{i,j-1}+\delta_{i,j+1}) 
 \end{equation}
describes diffusive coupling, and
 \begin{equation}
 N^t_{ij}=\partial_x f(\bar{x}^t_i,a)\delta_{i,j}
 \end{equation}
defines the strength of local instability, with $\delta_{i,j\pm 1}$ extended to
comply with periodic boundary condition. This partition of the Jacobian
explicitly shows how the symmetry group ${\mathcal L}$ depends on the symmetry
properties of the nonlinear evolution equation (\ref{eq_cml}) and those of the
controlled state $\bar{\bf x}^t$. The matrix $M$ has all the symmetries imposed
by the chosen inter-site couplings of the nonlinear model:
 \begin{equation}
 T(g)M=MT(g),\quad \forall g\in {\mathcal G},
 \end{equation}
while the matrix $N^t$ has all the symmetries of the target state $\bar{\bf
x}^t$: 
 \begin{equation}
 T(g)N^t=N^tT(g),\quad \forall g\in {\mathcal H}_{\bar{\bf x}},
 \end{equation}
where, $T$ denotes the matrix representation of ${\mathcal L}$, and ${\mathcal
H}_{\bar{\bf x}}$ is defined as a subgroup of ${\mathcal G}$. Since the
Jacobian $A^t$ commutes with all matrices that commute with both $M$ and $N^t$,
we conclude that generically ${\mathcal L}={\mathcal H}_{\bar{\bf x}}\subseteq
{\mathcal G}$, in agreement with the general result (\ref{eq_gt_prime}).

Since the symmetry analysis conducted above is valid for every subgroup
${\mathcal L}'$ of the dynamical symmetry group, we take ${\mathcal
L}'={\mathcal L}$. Constructing the $n_x$-dimensional representation $T$ of
${\mathcal L}$ and decomposing it into the sum of the irreducible
representations of ${\rm C}_{n_xv}$ we easily determine the restrictions
imposed by the symmetry on the minimal number of control parameters $n_u$ and
the structure of the control matrix $B$. For instance, a zigzag state gives
${\mathcal L}={\rm C}_{nv}$ with $n=n_x/2$ and, according to
(\ref{eq_max_blk}), $\bar{n}_u=2$; a non-reflection-invariant state with
spatial period $s$  corresponds to ${\mathcal L}={\rm C}_n$ with $n=n_x/s$ and
$\bar{n}_u=1$, etc. 

Let us consider the uniform target state, which has the highest symmetry
possible, ${\mathcal L}={\rm C}_{n_xv}$, in more detail. The decomposition
(\ref{eq_decomp}) gives
 \begin{equation}
 \label{eq_cnv_dec}
 T=T^1\oplus T^4\oplus T^5\oplus \cdots \oplus T^{n_x/2+3},
 \end{equation}
and the corresponding basis of normal modes which transform according to these
irreducible representations is given by the eigenvectors of the operators of
translation and reflection, i.e., Fourier modes
${\bf g}^i$:
 \begin{equation}
 \label{eq_mode}
 ({\bf g}^i)_j=\cos(jk_i+\phi_i).
 \end{equation}
Here $\phi_i$ are arbitrary phase shifts, and $k_i$ are the wavevectors defined
thus: $k_1=0$, $k_i=k_{i+1}=\pi i/n_x$ for $i=2,4,6,\cdots$, and, for $n_x$ --
even, $k_{n_x}=\pi$. Fourier modes with the same wavevectors $k$ define
invariant subspaces $L^k\subset{\mathcal T}={\mathbb R}^{n_x}$. The subspaces
$L^k$ with $0<k<\pi$ correspond to the representations $T^r$ with $r\ge 5$,
$L^0$ corresponds to $T^1$, and $L^\pi$ to $T^4$. Since the two-dimensional
irreducible representations are present in the decomposition
(\ref{eq_cnv_dec}), $\bar{n}_u=2$. Therefore, in order to control an unstable
uniform steady state of the coupled map lattice we need at least two control
parameters. This is the reflection of symmetric coupling in the model
(\ref{eq_cml}). Note that, since every two-dimensional irreducible
representation occurs in the decomposition (\ref{eq_cnv_dec}) once,
$p_5=\cdots=p_{n_x/2+3}=1$, according to the results of section \ref{s_ltv},
the minimal number of control parameters remains the same for a spatially
uniform target trajectory of arbitrary time period $\tau$.

On the other hand, since for any length $n_x$ of the lattice the group
${\mathcal G}={\rm C}_{n_xv}$ only has one- and two-dimensional irreducible
representations and ${\mathcal L}$ is a subgroup of ${\mathcal G}$, it is
sufficient to have just two control parameters to make the dynamics of the
coupled map lattice controllable in the vicinity of a target state with
arbitrary symmetry properties and temporal period. Choosing the minimal number
of control parameters, $n_u=2$, we can determine the conditions making them
independent with respect to a particular target state: the linear response of
the CML to perturbation of the two parameters, given by the columns of the
control matrix $B=\left[\matrix{{\bf b}_1& {\bf b}_2}\right]$, has to satisfy
conditions (\ref{eq_invar_b}) and (\ref{eq_rank_cnd}).

Failure to satisfy the necessary condition (\ref{eq_invar_b}) rules out the
possibility of using global parameters, such as the coupling $\epsilon$ or
parameter $a$ of the local map $f(x,a)$ for control of symmetric steady states.
Taking ${\bf u}=(a,\epsilon)$, so that
 \begin{equation}
 \label{eq_resp_a}
 {\bf b}_1=\partial_a{\bf F}(\bar{\bf x},{\bf 0},\bar{\bf u})
 =M\left[\matrix{\partial_a f(\bar{x}_1,\bar{a})\cr
 \vdots\cr \partial_a f(\bar{x}_{n_x},\bar{a})}\right],
 \end{equation}
and
 \begin{equation}
 \label{eq_resp_e}
 {\bf b}_2=\partial_\epsilon {\bf F}(\bar{\bf x},{\bf 0},\bar{\bf u})
 =(\bar{\epsilon})^{-1}(M-I)\left[\matrix{f(\bar{x}_1,\bar{a})\cr
 \vdots\cr f(\bar{x}_{n_x},\bar{a})}\right],
 \end{equation}
we observe that condition (\ref{eq_invar_b}) is only satisfied, if the group
${\mathcal L}$ is trivial, ${\mathcal L}=\{e\}$. This result holds for
time-periodic symmetric target states as well.

Alternatively, one can make the system controllable by directly perturbing the
system at the sites $i_1$ and $i_2$. The positions of the ``control'' sites
cannot be chosen arbitrary, again due to symmetry. However, if the target state
is spatially uniform, it is trivial to show that choosing, e.g., $i_1=l$ and
$i_2=l+1$ satisfies the controllability condition for an arbitrary length of
the lattice $n_x$. The control matrix corresponding to this choice of control
parameters can be written in the form
$B_{ij}=\delta_{j,1}\delta_{i,l}+\delta_{j,2}\delta_{i,l+1}$.

Such localized control also has its downside. In the weak coupling limit,
$\epsilon \rightarrow 0$, the coupled map lattice with local feedback becomes a
weakly controllable system. The symmetry of the lattice of uncoupled maps is
described by the permutation group ${\mathcal G}={\rm S}_{n_x}$, while the
linearization about a uniform target state increases the symmetry to ${\mathcal
L}={\rm GL}(n_x)$: since the respective Jacobian is a multiple of the unit
matrix, $A_{ij}=\partial_x f(\bar{x},\bar{a}) \delta_{i,j}$, the linearized
system is symmetric with respect to all (complex) nonsingular coordinate
transformations. When coupling is restored, $\epsilon>0$, the symmetry of both
the nonlinear evolution equation (\ref{eq_cml}) and its linearization
(\ref{eq_lin_id}) reduces to ${\mathcal G}'={\mathcal L}'={\rm C}_{n_xv}$.

The matrix representation $T$ of the group ${\rm GL}(n_x)$ in ${\mathcal T}$ is
already irreducible. Consequently, $n_u=n_x$ independent control parameters are
required to control the steady uniform state of the uncoupled lattice. This
result is rather intuitive. Obviously, one can no longer control the system
applying control perturbations at just two lattice sites, $i_1$ and $i_2$.
Since the control perturbation does not propagate to adjacent sites of the
lattice, feedback has to be applied directly at each site.

If the coupling is nonzero, but very small, the controllability property is
restored for $n_u=2$, but, according to section \ref{s_violat}, feedback of
very large magnitude is required to control the system due to parametric
deficiency. Indeed, in order to affect the dynamics at site $i$ away from $i_1$
and $i_2$ the control has to propagate a certain distance decaying by roughly a
factor of $\epsilon$ per iteration. As a result, the magnitude of the
perturbation required to control an arbitrary site of the lattice diverges
approximately as $\epsilon^{-n_x/2}$ for $\epsilon\rightarrow 0$, resulting in
the loss of control \cite{self_prl}.

%%%%%%%%%%%%%%%%%%%%%%%%%%%%%%%%%%%%%%%%%%%%%%%%%%%%%%%%%%%%%%%%%%%%%%%%%%%%%%%
\subsection{Particle in a Symmetric Potential}
\label{ss_particle}
%%%%%%%%%%%%%%%%%%%%%%%%%%%%%%%%%%%%%%%%%%%%%%%%%%%%%%%%%%%%%%%%%%%%%%%%%%%%%%%

The motion of a particle in a symmetric potential, such as a point charge in
electric field, serves as another example of the relation between the
structural symmetry group ${\mathcal G}$ and the dynamical symmetry group
${\mathcal L}$. This and many other interesting physical systems, e.g.,
inverted pendulum, or a satellite in orbit, are described by the second order
ordinary differential equation
 \begin{equation}
 \label{eq_ham}
 m\ddot{\bf r}=-\nabla V({\bf r}),
 \end{equation}
which can be trivially reduced to a system of first order differential
equations of the form (\ref{eq_nl_id_cont}) introducing additional coordinate
${\bf v}=\dot{\bf r}$. Suppose the potential $V({\bf r})$ possesses the cubic
symmetry (described by the group ${\rm O}$ which is a subgroup of ${\rm
SO(3)}$), but is not spherically symmetric, for instance:
 \begin{equation}
 V({\bf r})=V_0\cosh(kx)\cosh(ky)\cosh(kz).
 \end{equation}
The group ${\mathcal G}={\rm O}$ defines the structural symmetry of the
evolution equation (\ref{eq_ham}). Linearizing the latter about the equilibrium
position $\bar{\bf r}=0$ we obtain
 \begin{eqnarray}
 \label{eq_ham_lin}
 \partial_t\left[\matrix{{\bf r}\cr{\bf v}}\right]=
 \left[\matrix{0 & I \cr \omega^2 I & 0}\right]
 \left[\matrix{{\bf r}\cr{\bf v}}\right],
 \end{eqnarray}
where $\omega^2=-V_0 k^2/m$, while $0$ and $I$ are $3\times 3$ zero and unit
blocks, respectively. If $V_0<0$ the equilibrium is unstable, and control
should be applied to keep the system close to the equilibrium state.  

Equation (\ref{eq_ham_lin}) is spherically symmetric, with ${\mathcal L}'={\rm
SO(3)}$ and, therefore, ${\mathcal G}\subset {\mathcal L}$, i.e., the symmetry
of the linearized equation is higher than the symmetry of the original
nonlinear evolution equation. (In fact, the full symmetry group of equation
(\ref{eq_ham_lin}) is ${\mathcal L}={\rm GL}(3)$, but we choose to use its
subgroup ${\mathcal L}'={\rm SO}(3)$, since it is physically more relevant,
completely resolves the structure of the Jacobian matrix and, as such,
correctly represents the effect of symmetry on the control setup.)

Next we notice that the representation $T$ of the group ${\mathcal L}'$ in the
six-dimensional tangent space $\{{\bf r},{\bf v}\}$ can be decomposed into a
sum of two equivalent three-dimensional irreducible representations of ${\rm
SO(3)}$ (vector representations, which coincide with the respective irreducible
representation of ${\rm GL}(3)$):
 \begin{equation}
 T=2T^1,\qquad d_1=3.
 \end{equation}
This indicates that in order to control the unstable steady state $\bar{\bf r}
=\bar{\bf v}=0$ one needs at least three independent control parameters,
$\bar{n}_u=3$.

Arguably the simplest way to control such a system is to re-adjust the
potential (applying external fields, shifting support point, etc.) based
on the instantaneous values of the position ${\bf r}$ and velocity ${\bf
v}$ of the particle. This corresponds to picking the control matrix in the
following form:
 \begin{equation}
 B=\left[\matrix{ 0 & 0 & 0 \cr {\bf b}_1 & {\bf b}_2 & {\bf b}_3}\right],
 \end{equation}
 where ${\bf b}_1$, ${\bf b}_2$, ${\bf b}_3$ could be chosen as any three
linearly independent vectors in ${\mathbb R}^3$.

%%%%%%%%%%%%%%%%%%%%%%%%%%%%%%%%%%%%%%%%%%%%%%%%%%%%%%%%%%%%%%%%%%%%%%%%%%%%%%%
\section{Conclusions}
\label{s_summary}
%%%%%%%%%%%%%%%%%%%%%%%%%%%%%%%%%%%%%%%%%%%%%%%%%%%%%%%%%%%%%%%%%%%%%%%%%%%%%%%

Summarizing, we have determined that if the system under consideration is
symmetric, it cannot be considered generic with respect to conventional chaos
control techniques, and its symmetry properties should be understood prior to
constructing a control scheme, even if the symmetry is only approximate. The
failure to observe the restrictions imposed by the symmetry on the structure of
the measured output signal will usually prevent the experimental reconstruction
of the system dynamics. Similarly, an inappropriate choice of control
parameters will result in weak controllability and, as a result, extreme
sensitivity to noise, or even worse, complete loss of control.

From the practical point of view, the main result of the symmetry analysis is
that the minimal number of independent control parameters required for control,
as well as the minimal number of independent scalar observables required for
the reconstruction of local dynamics, can typically be determined without any
knowledge of the evolution equations governing the dynamics of the system. One
only needs to know the symmetry properties, such as spatial and temporal
periodicity, of the target state, and the structural symmetry of the dynamical
equations, which in the case of extended chaotic systems is often uniquely
defined by the geometry of the underlying physical space. One should, however,
realize that this typical pattern does not apply to all symmetric systems
without exception. The dynamical equations might, in principle, be symmetric
with respect to transformations unrelated to ``geometrical'' symmetries, such
as rotational, reflectional, or translational invariance. Additional
``nonphysical'' symmetries can also be introduced as a result of the
linearization procedure.

A number of comments have to be made regarding accidental degeneracies. We
found that when accidental degeneracies are present, restrictions obtained
using symmetry considerations alone provide only the necessary conditions for
controllability. In particular, one obtains a lower bound on the minimal number
of control parameters. Exact determination of that number in this case requires
additional information about the structure of the Jacobian matrix, which can be
gathered using experimental reconstruction. On the other hand, experimental
reconstruction itself is only possible, if there is an adequate number of
independent scalar observables. This number, however, is similarly
undetermined. In practice, though, one rarely has to worry about such
complications, since accidental degeneracies are not common and  unlikely to be
a problem for most actual experimental systems. Besides, an estimate for the
minimal number of observables and control parameters can always be easily
obtained using combinatorial arguments. Also, one should be careful in equating
the minimal number of observables or control parameters with the highest
degeneracy of the Jacobian matrix, especially if this degeneracy is at least
partially accidental. It can be argued that accidental degeneracies between
eigenvalues from the same irreducible invariant subspace typically will not
increase the dimensionality of the respective eigenspace and, therefore will
not lead to additional degeneracy in the local dynamics.

We also established that it is not enough to find an adequate number of control
parameters (or observables). These control parameters (observables) have to
satisfy certain conditions. In particular, perturbation of the control
parameters should completely break the dynamical symmetry. The more strict
independence condition is specific to each target trajectory and, on the one
hand, requires the knowledge of the system's response to variation of different
control parameters (which can be obtained experimentally, if necessary), but,
on the other hand, allows one to choose the minimal set of control parameters
systematically, avoiding trial and error search. For example, in case of
extended dynamical systems with local feedback the independence condition
usually imposes restrictions on the mutual arrangement of ``control'' sites,
while the number of ``control'' sites coincides with the number of control
parameters.

Finally, we discovered that the conventional approach to system identification
also has to be modified in the presence of symmetries. In particular, in order
to preserve not only the topology of the original attractor, but also the
symmetry of the original dynamical equations, one has to use a number of
simultaneously measured observables, which have to be the components of an
equivariant vector function of the actual state of the system. The restrictions
on the output can be relaxed somewhat in the case of local reconstruction in
the vicinity of some target trajectory. However, even then a number of
independent observables should be used instead of just a single one, as long as
the symmetry of the target state is nontrivial, leading to the increase in the
dimension of the embedding space. Otherwise, the conventional approach carries
over with minor modifications.

\end{document}